\pdfoutput=1
\documentclass[a4paper,12pt]{scrartcl}

\usepackage[utf8]{inputenc}
\usepackage[T1]{fontenc}

\usepackage[british]{babel}
\usepackage{csquotes}
\usepackage[
    doi=false,isbn=false,url=false,
    style=apa,
    backend=biber,
    uniquename=false,
    apamaxprtauth=18,
    sortcites=false]{biblatex}
\DeclareLanguageMapping{british}{british-apa}
\DeclareLabeldate{%
                 \field{year}
                 \field{date}
                 \field{eventdate}
                 \field{origdate}
                 \field{urldate}
                 \literal{n.d.}
}
\usepackage{libertine}
\usepackage[libertine]{newtxmath}
\usepackage{setspace}
\usepackage{xcolor}

\usepackage{xfrac}
\usepackage{mathtools}
\mathtoolsset{centercolon}

\usepackage{amssymb}
\usepackage{centernot}

\usepackage{xspace}
\newcommand{\ocmlong}{OraCle Modelling\xspace}
\newcommand{\ocm}{OCM\xspace}

\newcommand{\independent}{\mbox{${}\perp\mkern-11mu\perp{}$}}
\newcommand{\nindependent}{\centernot\independent}

\newcommand\pa[1]{{\ensuremath \mathbf{PA}(#1)}}
\newcommand\doop[1]{{\ensuremath \operatorname{do}(#1)}}

\usepackage{graphicx}
\usepackage{caption}
\usepackage{subcaption}
\usepackage{hyperref}
\definecolor{linkcolor}{HTML}{449B52}
\definecolor{citecolor}{HTML}{67001F}
\definecolor{urlcolor}{HTML}{008080}
\hypersetup{
    colorlinks,
    linkcolor={linkcolor},
    citecolor={citecolor},
    urlcolor={urlcolor}
}

\usepackage{tikz}
\graphicspath{{./graphics/}}
\usetikzlibrary{arrows.meta}
\usetikzlibrary{backgrounds}
\usetikzlibrary{decorations.pathreplacing}

\tikzset{node/.style={circle, draw, minimum size=2.5em, text centered, thick, fill=white}}
\tikzset{nodec/.style={circle, draw, minimum size=2.5em, text centered, thick, fill=lightgray}}
\tikzset{nodeh/.style={circle, dashed, draw, minimum size=2.5em, text centered, thick, fill=white}}
\tikzset{la/.style={-{Latex[length=4mm, width=4mm]},thick}}

\newcommand{\drawdatabox}[6]{%
\foreach \i in {1,...,5}
{
    \node[draw,rounded corners,minimum height=1.5cm,minimum width=1.5cm,draw=#3,fill=#4] at (#1 + .5 - \i * .1,#2 + .275 - \i * .055) {};
}
\node[draw,rounded corners,minimum height=1.5cm,minimum width=1.5cm,draw=#3,fill=#4] at (#1,#2) {#5};
\node[gray] at (#1 + 1.05,#2 - .68) {\tiny #6};
}

\author{Sebastian Weichwald \& Jonas Peters}
\title{%
Causality in cognitive neuroscience: concepts, challenges, and distributional robustness%
}
\date{July 2020}

\begin{document}

\maketitle

\begin{abstract}
    While probabilistic models
    describe the dependence structure between observed variables, causal models go one step further: they predict, for example,
    how cognitive functions are affected by external interventions that perturb neuronal activity.
In this review and perspective article, we introduce the concept of causality in the context of cognitive neuroscience and review existing methods for inferring causal relationships from data.
Causal inference is an ambitious task that is particularly challenging in cognitive neuroscience.
We
discuss two difficulties in more detail:
    the scarcity of interventional data and the challenge of
    finding the right variables.
    We argue for distributional robustness as a guiding principle to tackle these problems.
Robustness (or invariance)
is a fundamental principle underlying causal
methodology.
A causal model of a target variable
generalises
across environments or subjects
as long as these environments leave the causal mechanisms intact.
Consequently,
if a candidate model
does not generalise, then
either
it
does not consist of the target variable's causes
 or
the underlying variables
do not represent the correct granularity of the
problem.
In this sense, assessing
generalisability
may
be useful
when defining
relevant variables and can be used to partially compensate for the lack of interventional data.
\end{abstract}

\section{Introduction}\label{sec:introduction}

Cognitive neuroscience aims to describe and understand the neuronal underpinnings of cognitive functions such as perception, attention, or learning.
The objective is to characterise brain activity and cognitive functions, and to relate one to the other.
The submission guidelines for the Journal of Cognitive Neuroscience, for example, state:
``The Journal will not publish research reports that bear solely on descriptions of function without addressing the underlying brain events, or that deal solely with descriptions of neurophysiology or neuroanatomy without regard to function.''
We think that understanding this relation requires us to relate brain events and cognitive function in terms of the cause-effect relationships that govern their interplay.
A causal model could, for example,
describe how cognitive functions are affected by external interventions that perturb neuronal activity (cf.\ Section~\ref{sec:whycausality}).
\textcite{reid2019advancing} argue that ``the ultimate phenomenon of theoretical interest in all FC [functional connectivity] research is understanding the causal interaction among neural entities''.

Causal inference in cognitive neuroscience is of great importance and perplexity.
This motivates our discussion of two pivotal challenges.
First, the scarcity of interventional data is problematic as several causal models may be equally compatible with the observed data while making conflicting predictions only about the effects of interventions~(cf.\ Section~\ref{sec:challenge1}).
Second, the ability to understand how neuronal activity gives rise to cognition depends on finding the right variables to represent the neuronal activity~(cf.\ Section~\ref{sec:challenge2}).
Our starting point is the well-known observation that
causal models of a target (or response) variable are distributionally robust and thus generalise across environments, subjects, and interventional shifts~\parencite{Haavelmo1944,Aldrich1989,Pearl2009}. %
Models
that do not generalise
are either based upon the wrong variables that do not represent causal entities or include variables that are not causes of the target variable.
We thus propose to pursue robust (or invariant) models.
That way, distributional robustness may
serve as a guiding principle towards a causal understanding of cognitive function and may help us tackle both challenges mentioned above.

\subsection{Running examples}\label{sec:runningexample}

We consider the following simplified examples.
Assume that the consumption of alcohol affects reaction times in a cognitive task.
In a randomised controlled trial we find that drinking alcoholic (versus non-alcoholic) beer results in slowed reaction times hereinafter.
Therefore, we may write `$\text{alcohol} \to \text{reaction time}$' and call alcohol a cause of reaction time and reaction time an effect of alcohol.
Intervening on
the cause results in a change in
the distribution of the effect. In our example, prohibiting the consumption of any alcoholic beers results in faster reaction times.

\begin{figure}
\begin{subfigure}{\textwidth}
\centering
\begin{tikzpicture}
    \node (X) at(0,0) {{\reflectbox{\includegraphics[keepaspectratio,width=7cm]{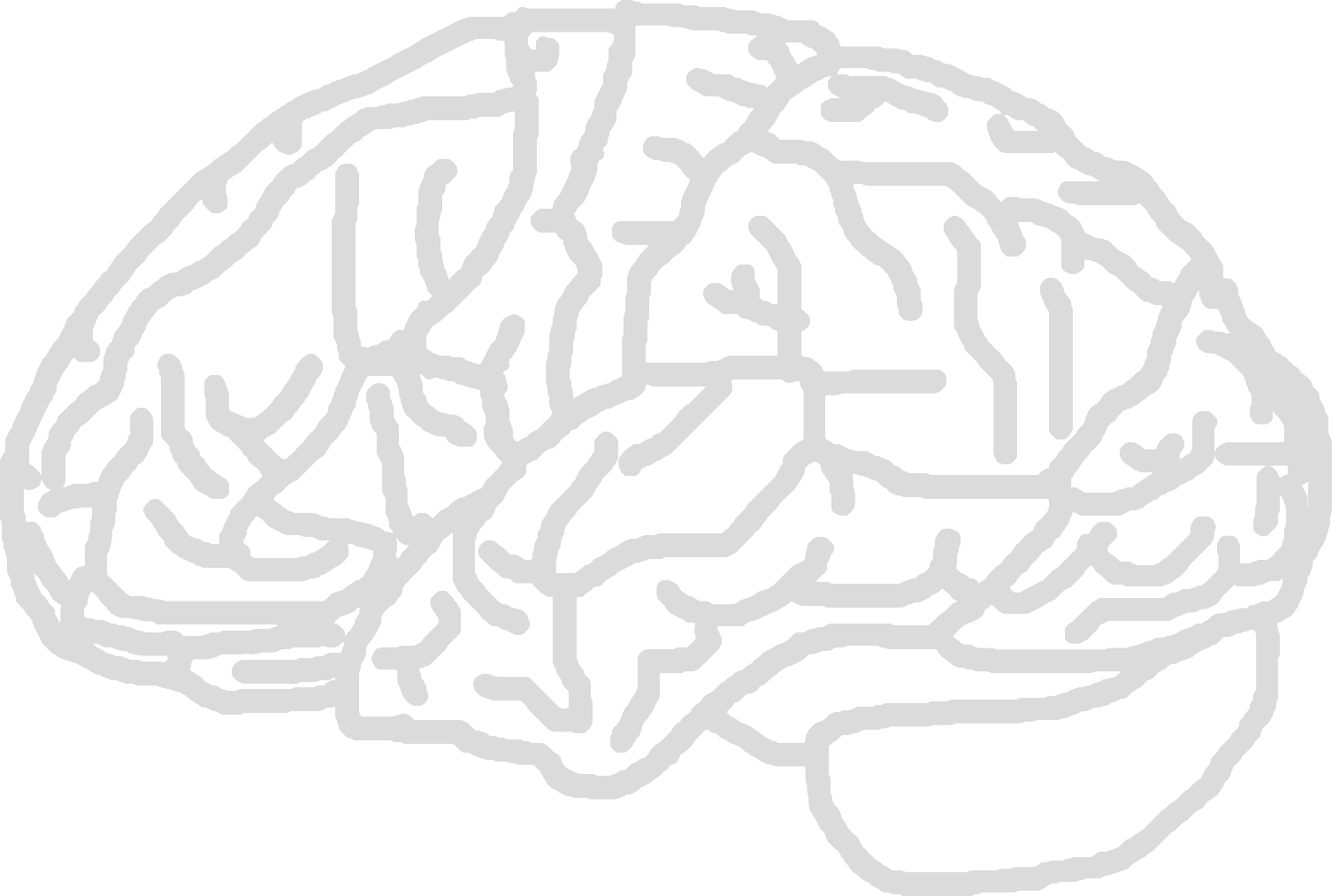}}}};

    \node[node] (T) at(-6,0) {$T$};
    \node[nodeh,draw=gray,text=gray] (H) at(2.5,3) {$H$};
    \node[node] (x1) at(-2,-1.5) {$X_1$};
    \node[node] (x2) at(0,1.5) {$X_2$};
    \node[node] (x3) at(2,0) {$X_3$};
    \node[node] (Y) at(6,0) {$Y$};

    \draw[->,la] (T) -- (x1);
    \draw[->,la] (T) -- (x2);
    \draw[->,la] (T) -- (x3);
    \draw[->,la,gray] (H) -- (x1);
    \draw[->,la,gray] (H) -- (x2);
    \draw[->,la,gray] (H) -- (x3);
    \draw[->,la,gray] (H) -- (Y);
    \draw[->,la] (x1) -- (Y);
    \draw[->,la] (x2) -- (Y);
    \draw[->,la] (x3) -- (Y);
    \draw[->,la] (x1) -- (x2);
    \draw[->,la] (x2) -- (x3);
    \draw[->,la] (x3) -- (x1);
    \draw[->,la] (x3) -- (x2);
    \draw[->,la] (x2) -- (x1);
    \draw[->,la] (x1) -- (x3);

    \node[fill=white,circle,text=red] (q1) at(-4.5,0) {{\LARGE ?}};
    \node[fill=white,circle,text=red] (q2) at(0,0) {{\LARGE ?}};
    \node[fill=white,circle,text=red] (q3) at(4.5,0) {{\LARGE ?}};
\end{tikzpicture}
\caption{Running Example A}\label{fig:toyexampleA}
\end{subfigure}

\begin{subfigure}{\textwidth}
\centering\hspace{-7em}
\begin{tikzpicture}
    \node (X) at(0,0) {{\reflectbox{\includegraphics[keepaspectratio,width=7cm]{brain.png}}}};

    \node[node] (T) at(-6,0) {$T$};
    \node[nodeh,draw=gray,text=gray] (H) at(2.5,3) {$H$};
    \node[node] (x1) at(-2,-1.5) {$X_1$};
    \node[node] (x2) at(0,1.5) {$X_2$};
    \node[node] (x3) at(2,0) {$Y$};

    \draw[->,la] (T) -- (x1);
    \draw[->,la] (T) -- (x2);
    \draw[->,la] (T) -- (x3);
    \draw[->,la,gray] (H) -- (x1);
    \draw[->,la,gray] (H) -- (x2);
    \draw[->,la,gray] (H) -- (x3);
    \draw[->,la] (x1) -- (x2);
    \draw[->,la] (x2) -- (x3);
    \draw[->,la] (x3) -- (x1);
    \draw[->,la] (x3) -- (x2);
    \draw[->,la] (x2) -- (x1);
    \draw[->,la] (x1) -- (x3);

    \node[fill=white,circle,text=red] (q1) at(-4.5,0) {{\LARGE ?}};
    \node[fill=white,circle,text=red] (q2) at(0,0) {{\LARGE ?}};
\end{tikzpicture}
\caption{Running Example B}\label{fig:toyexampleB}
\end{subfigure}

\caption{
Illustration of two scenarios in cognitive neuroscience where we seek a causal explanation focusing on a target variable $Y$ that either resembles (a) a cognitive function, or (b) a neuronal entity.
The variables $T$, $\mathbf{X}=[X_1,\dots,X_d]^\top$, and $H$ represent treatment, measurements of neuronal activity, and an unobserved variable, respectively.}\label{fig:toyexample}
\end{figure}

In cognitive neuroscience one may wish to describe how the neuronal activity is altered upon beer consumption and how this change in turn affects the reaction time.
For this, we additionally require a measurement of neuronal activity, say a functional magnetic resonance imaging (fMRI) scan and voxel-wise blood-oxygen-level dependent (BOLD) signals, that can serve as explanans in a description of the phenomenon
`$\text{alcohol} \to \text{neuronal activity} \to \text{reaction time}$'.
We distinguish the following two scenarios:

\paragraph{Running Example A, illustrated in Figure~\ref{fig:toyexampleA}.}
A so-called treatment or stimulus variable $T$ (say, consumption of alcohol) affects neuronal activity as measured by a $d$-dimensional feature vector $\mathbf{X} = [X_1,\dots,X_d]^\top$ and the target variable $Y$ reflects a cognitive function (say, reaction time).
We may concisely write $T \to \mathbf{X} \to Y$ for a treatment that affects neuronal activity which in turn maintains a cognitive function~\parencite[this is analogous to the `stimulus $\to$ brain activity $\to$ response' set-up considered in][]{weichwald2014causal,weichwald2015causal}.

\paragraph{Running Example B, illustrated in Figure~\ref{fig:toyexampleB}.}
We may wish to describe how neuronal entities cause one another and hence designate one such entity as the target variable $Y$.
In this example, we consider a target variable corresponding to a specific brain signal or region instead of a behavioural or cognitive response.

\subsection{Existing work on causality in the context of cognitive neuroscience}\label{sec:existingwork}
Several methods
such as Granger causality or constraint-based methods
have been applied to the problem of
inferring causality from cognitive neuroscience data.
We describe these methods
in Section~\ref{sec:discovery}.
In addition, there are
ongoing conceptual debates
that revolve around the principle of causality in cognitive neuroscience, some of which
we now mention.
\textcite{mehler2018the} raise concerns about the ``lure of causal statements'' and expound the problem of confounding when interpreting functional connectivity.
Confounders are similarly problematic for multi-voxel pattern analyses~\parencite{todd2013confounds,woolgar2014coping}.
The causal interpretation of encoding and decoding (forward and backward, univariate and multivariate) models has received much attention as they are common in the analysis of neuroimaging data:
\textcite{davis2014what} examine the differences between the model types, \textcite{haufe2014on} point out that the weights of linear backward models may be misleading, and \textcite{weichwald2015causal} extend the latter argument to non-linear models and clarify which causal interpretations are warranted from either model type.
Feature relevance in mass-univariate and multivariate models can be linked to marginal and conditional dependence statements that yield an enriched causal interpretation when both are combined~\parencite{weichwald2015causal}; this consideration yields refined results in neuroimaging analyses~\parencite{huth2016decoding,bach2017whole,varoquaux2018atlases} and explains improved functional connectivity results when combining bivariate and partial linear dependence measures~\parencite{sanchez2019combining}.
Problems such as indirect measurements and varying temporal delays complicate causal Bayesian network approaches for fMRI \parencite{ramsey2010six,mumford2014bayesian}.
\textcite{smith2011network} present a simulation study evaluating several methods for estimating brain networks from fMRI data and demonstrate that identifying the direction of network links is difficult.
The discourse on how to leverage connectivity analyses to understand mechanisms in brain networks is ongoing~\parencite{valdes2011effective,waldorp2011effective,smith2012future,mill2017connectome}.
Many of the above problems and findings are related to the two key challenges that we discuss in Section~\ref{sec:challenges}.

\subsection{Structure of this work}

We begin Section~\ref{sec:causality} by formally introducing causal concepts.
In Section~\ref{sec:whycausality}, we outline why we believe there is a need for causal models in cognitive neuroscience by considering what types of questions could be answered by an \ocmlong (\ocm) approach.
We discuss the problem of models that are observationally equivalent yet make conflicting predictions about the effects of interventions in Section~\ref{sec:modelequivalence}.
In Section~\ref{sec:discovery}, we review different causal discovery methods and their underlying assumptions.
We focus on two challenges for causality in cognitive neuroscience that are expounded in Section~\ref{sec:challenges}: (1) the scarcity of interventional data and
(2) the challenge of finding the right variables.
In Section~\ref{sec:robustness}, we argue that one should seek distributionally robust variable representations and models to tackle these challenges.
Most of our arguments in this work are presented in an i.i.d.\ setting
and we briefly discuss the implications for time-dependent data in Section~\ref{sec:furtherprobs}.
We conclude in Section~\ref{sec:conclusion} and outline ideas
that we regard as promising for future research.

\section{Causal models and causal discovery}\label{sec:causality}

In contrast to classical probabilistic models, causal models induce not only an observational distribution but also a set of so-called interventional distributions.
That is, they predict how a system reacts under interventions.
We present an introduction to causal models that is based on pioneer work by \textcite{Pearl2009} and \textcite{Spirtes2000}.
Our exposition is inspired by \textcite[Chapter 2]{weichwald2019pragmatism}, which provides more introductory intuition into causal models viewed as structured sets of interventional distributions.
For both simplicity and focus of exposition, we omit a discussion of counterfactual reasoning and other akin causality frameworks such as the potential outcomes formulation of causality~\parencite{imbens2015causal}.
We phrase this article within the framework and terminology of Structural Causal Models (SCMs)~\parencite{bollen1989structural,Pearl2009}.

An SCM over variables $\mathbf{Z} = [Z_1,\dots,Z_d]^\top$ consists of %
\begin{description}
 \item[structural equations] that relate each variable $Z_k$ to its parents $\pa{Z_k} \subseteq \{Z_1,\dots,Z_d\}$ and a noise variable $N_k$ via a function $f_k$ such that $Z_k := f_k(\pa{Z_k}, N_k)$, and a
 \item[noise distribution $\mathbf{P}_{\mathbf{N}}$] of the noise variables $\mathbf{N} = [N_1,\dots,N_d]^\top$.
\end{description}
We associate each SCM with a directed causal graph where the nodes correspond to the variables $Z_1,\dots,Z_d$ and we draw an edge from $Z_i$ to $Z_j$ whenever $Z_i$ appears on the right hand side of the equation $Z_j := f_j(\pa{Z_j},N_j)$.
That is, if $Z_i \in \pa{Z_j}$ the graph contains the edge $Z_i \to Z_j$.
Here, we assume that this graph is acyclic.
The structural equations and noise distributions together induce the observational distribution $\mathbf{P_Z}$ of $Z_1,\dots,Z_d$ as simultaneous solution to the equations.
(\textcite{bongers2018theoretical} formally define SCMs when the graph includes cycles.)

The following is an example of a linear Gaussian SCM:
\begin{align*}
Z_1 &:= f_1(\pa{Z_1},N_1) = Z_2 + N_1 \\
Z_2 &:= f_1(\pa{Z_2},N_2) = N_2 \\
Z_3 &:= f_3(\pa{Z_3},N_3) = Z_1 + 5\cdot Z_2 + N_3
\end{align*}
with mutually independent standard-normal noise variables $N_1,N_2,N_3$.
The corresponding graph is
\begin{center}
\begin{tikzpicture}
    \node (z1) at(0,0) {$Z_1$};
    \node (z2) at(-1,-1) {$Z_2$};
    \node (z3) at(1,-1) {$Z_3$};

    \draw[->] (z2) -- (z1);
    \draw[->] (z1) -- (z3);
    \draw[->] (z2) -- (z3);
\end{tikzpicture}
\end{center}
and the SCM induces the observational distribution $\mathbf{P_Z}$, which is the multivariate Gaussian distribution
\begin{equation} \label{eq:exampleSCM}
\begin{pmatrix}
 Z_1 \\ Z_2 \\ Z_3
\end{pmatrix}
\sim
\mathbf{P_Z}
=
\mathcal{N}
\left(
\begin{pmatrix} 0 \\ 0 \\ 0\end{pmatrix},
\begin{pmatrix}
2 & 1 & 7 \\
1 & 1 & 6 \\
7 & 6 & 38
\end{pmatrix}
\right)\text{.}
\end{equation}

In addition to the observational distribution, an SCM induces interventional distributions.
Each intervention denotes a scenario in which we fix a certain subset of the variables to a certain value.
For example, the intervention $\doop{Z_2:=0,Z_3:=5}$ denotes the scenario where we force $Z_2$ and $Z_3$ to take on the values $0$ and $5$, respectively.
The interventional distributions are obtained by (a)~replacing the structural equations of the intervened upon variables by the new assignment, and (b)~considering the distribution induced by the thus obtained new set of structural equations.
For example, the distribution under intervention $\doop{Z_1:=a}$ for $a\in\mathbb{R}$, denoted by $\mathbf{P_Z}^\doop{Z_1:=a}$, is obtained by changing the equation $Z_1 := f_1(\pa{Z_1},N_1)$ to $Z_1 := a$.
In the above example, we find
\[
\mathbf{P_Z}^\doop{Z_1:=a}
=
\mathcal{N}
\left(
\begin{pmatrix} a \\ 0 \\ a\end{pmatrix},
\begin{pmatrix}
0 & 0 & 0 \\
0 & 1 & 5 \\
0 & 5 & 26
\end{pmatrix}
\right)\text{,}
\]
where $X\sim \mathcal{N}(a, 0)$ if and only if $\mathbf{P}(X=a)=1$.
Analogously, for $b\in\mathbb{R}$ and intervention on $Z_2$ we have
\[
\mathbf{P_Z}^\doop{Z_2:=b}
=
\mathcal{N}
\left(
\begin{pmatrix} b \\ b \\ 6\cdot b\end{pmatrix},
\begin{pmatrix}
1 & 0 & 1 \\
0 & 0 & 0 \\
1 & 0 & 2
\end{pmatrix}
\right)\text{.}
\]

The distribution of $Z_1$ differs between the observational distribution and the interventional distribution, that is,
$\mathbf{P}_{Z_1} \neq \mathbf{P}_{Z_1}^\doop{Z_2:=b}$.
We call a variable $X$ an (indirect) cause of a variable $Y$ if there exists an intervention on $X$ under which the distribution of $Y$ is different from its distribution in the observational setting.
Thus, $Z_2$ is a cause of $Z_1$.
The edge $Z_2\to Z_1$ in the
above causal graph reflects this cause-effect relationship.
In contrast, $Z_2$ remains standard-normally distributed under all interventions $\doop{Z_1:=a}$ on $Z_1$.
Because the distribution of $Z_2$ remains unchanged under any intervention on $Z_1$, $Z_1$ is not a cause of $Z_2$.

In general, interventional distributions do not coincide with the corresponding conditional distributions.
In our example we have $\mathbf{P}_{\mathbf{Z}\mid Z_1=a} \neq \mathbf{P_Z}^\doop{Z_1:=a}$ while $\mathbf{P}_{\mathbf{Z}\mid Z_2=b} = \mathbf{P_Z}^\doop{Z_2:=b}$.
We further have that the conditional distribution $\mathbf{P}_{Z_3\mid Z_2,Z_1}$ of $Z_3$ given its parents $Z_1$ and $Z_2$ is invariant under interventions
on variables other than $Z_3$.
We call a model of $Z_3$ based on $Z_1, Z_2$ invariant (cf.\ Section~\ref{sec:robustcausalmodels}).

We have demonstrated how an SCM induces a set of observational and interventional distributions.
The interventional distributions predict observations of the system upon intervening on some of its variables.
As such, a causal model holds additional content compared to a common probabilistic model that amounts to one distribution to describe future observations of the same unchanged system.
Sometimes we are only interested in modelling certain interventions or cannot perform others as there may be no well-defined corresponding real-world implementation.
For example, we cannot intervene on a person's gender.
In these cases it may be helpful to explicitly restrict ourselves to a set of interventions of interest.
Furthermore, the choice of an intervention set puts constraints on the granularity of the model~\parencite*[cf.\ Section~\ref{sec:challenge2} and Rubenstein \& Weichwald et al.,][]{rubenstein2017causal}.

\subsection{When are causal models important?}\label{sec:whycausality}
We do not always need causal models to answer our research question.
For some scientific questions it suffices to consider probabilistic, that is, observational models.
For example, if we wish to develop an algorithm for early diagnosis of Alzheimer's disease from brain scans, we need to model the conditional distribution of Alzheimer's disease given brain activity.
Since this can be computed from the joint distribution, a probabilistic model suffices.
If, however, we wish to obtain an understanding that allows us to optimally prevent progression of Alzheimer's disease by, for example, cognitive training or brain stimulation, we are in fact interested in
a causal understanding of the Alzheimer's disease and require a causal model.

Distinguishing between these types of questions is important
as it informs us about the methods we need to employ in order to answer the question at hand.
To elaborate upon this distinction, we now discuss scenarios related to our running examples and the relationship between alcohol consumption and reaction time (cf.\ Section~\ref{sec:runningexample}).
Assume we have access to a powerful \ocmlong (\ocm) machinery that is unaffected by statistical problems such as model misspecification, multiple-testing, or small sample sizes.
By asking ourselves, what queries must be answered by \ocm for us to `understand' the cognitive function, the difference between causal and non-causal questions becomes apparent.

Assume, firstly, we ran the reaction task experiment with multiple subjects, fed all observations to our \ocm machinery, and have Kim visiting our lab today.
Since \ocm yields
us the exact conditional distribution of reaction times $\mathbf{P}_{Y\mid T=t}$ for Kim having consumed $T=t$ units of alcoholic beer,
we may be willing to bet against our colleagues on how Kim will perform in the reaction task experiment they are just about to participate in.
No causal model for brain activity is necessary.

Assume, secondly, that we
additionally record BOLD responses $\mathbf{X}=[X_1,\dots,X_d]^\top$ at certain locations and times during the reaction task experiment.
We can query \ocm for the distribution of BOLD signals that we are about to record, that is, $\mathbf{P}_{\mathbf{X}\mid T=t}$, or
the distribution of reaction times
given we measure Kim's BOLD responses $\mathbf{X}=\mathbf{x}$, that is, $\mathbf{P}_{Y\mid T=t,\mathbf{X}=\mathbf{x}}$.
As before, we may bet against our colleagues on how Kim's BOLD signals will look like in the upcoming reaction task experiment or bet on their reaction time once we observed the BOLD activity $\mathbf{X} = \mathbf{x}$ prior to a reaction cue. Again, no causal model for brain activity is required.

In both of the above situations, we have learned something useful.
Given that the data were obtained in an experiment
in which
alcohol consumption was randomised, we
have learned, in the first situation, to predict
reaction times
after an intervention on alcohol consumption.
This may be considered an operational model for alcohol consumption and reaction time.
In the second situation, we have learned how the BOLD signal responds to alcohol consumption.
Yet, in none of the above situations have we
gained
understanding of the neuronal underpinnings of the cognitive function and the reaction times.
Knowing the conditional distributions $\mathbf{P}_{Y\mid T=t}$ and
$\mathbf{P}_{Y\mid T=t,\mathbf{X}=\mathbf{x}}$
for any $t$
yields no insight into any of the following
questions.
Which brain regions maintain fast reaction times?
Where in the brain should we release drugs that excite neuronal activity in order to counterbalance the effect of alcohol?
How do we need to update our prediction if we learnt that Kim just took a new drug that lowers blood pressure in the prefrontal cortex?
To answer such questions, we require causal understanding.

If we had a causal model, say in form of an SCM,
we could address the above questions.
An SCM offers an explicit way to
model the system under manipulations.
Therefore, a causal model can help to
answer questions about where to release an excitatory drug.
It may enable us to predict whether
medication that lowers blood pressure in the prefrontal cortex will affect Kim's reaction time;
in general,
this is the case if the corresponding variables appear
in the structural equations for $Y$ or any of $Y$'s ancestors.

Instead of identifying conditional distributions, one may formulate the problem as a regression task with the aim to learn the conditional mean functions
$t\to \mathbb{E}[\mathbf{X}|T=t]$ and $(t,\mathbf{x}) \to \mathbb{E}[Y|T=t, \mathbf{X} = \mathbf{x}]$.
These functions
are then parameterised in terms of $t$ or $t$ and
$\mathbf{x}$.
We argue in Section~\ref{sec:modelequivalence}, point (2),
that such parameters do not carry a causal meaning and thus do not help to answer the questions above.

Promoted by slogans such as `correlation does not imply causation' careful and associational language is sometimes used in the presentation of cognitive neuroscience studies.
We believe, however, that a clear language
that states whether a model should be interpreted causally (that is, as an interventional model) or
non-causally (that is, as an observational model) is needed.
This will help to clarify both the real world processes the model can be used for and the purported scientific claims.

Furthermore,
causal models may generalise better than non-causal models.
We expect
systematic
differences
between subjects and
between different trials or recording days of the same
subject.
These
different
situations,
or environments,
are presumably not arbitrarily different.
If they were,
we could not hope to gain
any scientific insight from
such experiments.
The apparent question is, which
parts of the model we can expect
to generalise between environments.
It is well-known that
causal models
capture one such invariance property,
which is implicit in the definition
of interventions.
An intervention on one variable leaves the
assignments of the other variables unaffected.
Therefore, the conditional distributions of
these other variables, given their parents, are also unaffected by the
intervention \parencite{Haavelmo1944, Aldrich1989}.
Thus,
causal models may enable us to formulate more clearly
which mechanisms we assume to be invariant between subjects. %
For example, we may assume that
the mechanism how alcohol intake affects
brain activity differs between subjects, whereas the mechanism
from signals in certain brain regions to reaction time is invariant.
We discuss the connection between causality and
robustness
in Section~\ref{sec:robustness}.

\subsection{Equivalences of models}\label{sec:modelequivalence}
Causal models entail strictly more information than observational models.
We now introduce the notion of equivalence of models~\parencite{Pearl2009,peters2017elements,bongers2018theoretical}.
This notion
allows us to discuss
the falsifiability of
causal models, which is important when
assessing candidate models and
their ability to
capture cause-effect relationships that govern a cognitive process under investigation.

We call two models observationally equivalent if they induce the same observational distribution.
Two models are said to be interventionally equivalent if they induce the same observational and interventional distributions.
As discussed above, for some interventions there may not be a
well-defined corresponding experiment in the real world.
We therefore also consider interventional equivalence
with respect to a restricted set of interventions.

One reason why learning causal models from
observational data is difficult is
the existence of
models that are
observationally %
but
not interventionally equivalent.
Such models agree in their predictions about the observed system yet
disagree in their predictions about the effects of certain interventions.
We continue the example from Section~\ref{sec:causality} and consider the following two SCMs:
\begin{align*}
 Z_1 &:= Z_2 + N_1
 &{Z}_1 &:= \sqrt{2}\cdot N_1 \\
 Z_2 &:= N_2
 &{Z}_2 &:=  \sfrac{1}{2}\cdot {Z}_1 + \sfrac{1}{\sqrt{2}}\cdot N_2 \\
 Z_3 &:= Z_1 + 5 \cdot Z_2 + N_3
 &{Z}_3 &:= {Z}_1 + 5\cdot {Z}_2 + N_3
\end{align*}
where in both cases $N_1,N_2,N_3$ are mutually independent standard-normal noise variables.
The two SCMs are observationally equivalent as they induce the same observational distribution, the one shown in Equation~\eqref{eq:exampleSCM}.
The models are not interventionally equivalent, however,
since
$\mathbf{P}_{Z_2}^\doop{Z_1:=3} = \mathcal{N}(0,1)$
and
$\mathbf{P}_{Z_2}^\doop{Z_1:=3} = \mathcal{N}(\sfrac{3}{2},\sfrac{1}{2})$
for the left and right model, respectively.
The two models can be told apart when interventions on
$Z_1$ or $Z_2$ are considered.
They are interventionally equivalent with respect to interventions on $Z_3$.

The existence of observationally equivalent models that are not interventionally
equivalent
has several implications.
(1)~Without assumptions,
it is impossible to learn causal structure from observational data.
This is not exclusive to causal inference from data and an analogous statement holds true for regression~\parencite{Gyoerfi2002}.
The regression problem is solvable only under certain simplicity assumptions, for example, on the smoothness of the regression function, which have been proven useful in real world applications.
Similarly,
there are several assumptions that can be exploited for causal discovery.
We discuss some of these assumptions in Section~\ref{sec:discovery}.
(2)~As a consequence, without further restrictive assumptions on the data generating process,
the estimated parameters do not carry any causal meaning.
For example, given any finite sample from the observational distribution,
both of the above SCMs yield exactly the same likelihood.
Therefore,
the above structures
cannot be told apart by a method that employs the maximum likelihood estimation principle.
Instead, which SCM and thus which parameters are selected in such a situation may depend on starting values, optimisation technique, or numerical precision.
(3)~Assume that we are given a probabilistic (observational) model
of a data generating process.
To falsify it, we may
apply a goodness-of-fit test based on an observational sample from that process.
An interventional model
cannot be falsified based on observational data alone and one has to also take into account the outcome of interventional experiments.
This requires that we are in agreement about how to perform the intervention in practice (see also Section~\ref{sec:challenge2}).
Interventional data may be crucial in particular for rejecting some of the observationally equivalent models (cf.\ the example above).
The scarcity of interventional data therefore poses a challenge for causality in cognitive neuroscience (cf.\ Section~\ref{sec:challenge1}).

\subsection{Causal discovery}\label{sec:discovery}
The task of learning a causal model from observational (or a combination of observational and interventional) data is commonly referred to as causal discovery or causal structure learning.
We have argued in the preceding section that causal discovery from purely observational data
is impossible without any additional assumptions or background knowledge.
In this section,
we discuss several assumptions that render (parts of) the causal structure identifiable from the observational distribution.
In short, assumptions concern how causal links manifest
in observable statistical dependences, functional forms of the mechanisms,
certain invariances under interventions, or
the order of time.
We briefly outline how these assumptions can be exploited in algorithms.
Depending on the application at hand, one may be
interested in learning the full causal structure as represented by its graph or in identifying a local structure such as the
causes of a target variable $Y$.
The methods described below
cover either of the two cases.
We keep the description brief focussing on the main ideas and intuition,
while more details can be found in the respective references.

\paragraph{Randomisation.}
The often called `gold standard' to establishing whether $T$ causes $Y$ is to introduce
controlled
perturbations, that is, targeted interventions, to a system.
Without randomisation, a dependence between
$T$ and $Y$ could stem from a confounder
between $T$ and $Y$ or from a causal link from $Y$ to $T$.
If $T$ is randomised
it is no further governed by the outcome of any other variable or mechanism.
Instead, it only depends on the outcome of a randomisation experiment, such as the roll of a die.
If
we observe that under the randomisation,
$Y$ depends on $T$, say the higher $T$ the higher $Y$, then there must be a (possibly indirect) causal influence from $T$ to $Y$.
In our running examples, this allows us to conclude that the amount of alcoholic beer consumed causes reaction times~(cf.\ Section~\ref{sec:runningexample}).
When falsifying interventional models, it
suffices to consider
randomised experiments as interventions \parencite[][Proposition~6.48]{peters2017elements}.
In practice, however, performing randomised experiments is often infeasible due to cost or ethical concerns, or impossible as, for example, we cannot randomise gender nor fully control neuronal activity in the temporal lobe.
While it is sometimes argued that the
experiment conducted by James Lind in 1747 to identify a treatment for scurvy
is among the first
randomised controlled trials,
the
mathematical theory and
methodology was popularised by Ronald A.\ Fisher in the early 20th century~\parencite{conniffe1991ra}.

\paragraph{Constraint-based methods.}
Constraint-based methods rely on two assumptions that connect
properties of the causal graph with
conditional independence statements in the induced distribution.
The essence of the first assumption is sometimes described as Reichenbach's common cause  principle~\parencite{reichenbach1956the}:
If $X$ and $Y$ are dependent, then there must be
some cause-effect structure that explains the observed dependence, that is, either $X$ causes $Y$, or $Y$ causes $X$, or another unobserved variable $H$ causes both $X$ and $Y$, or some combination of the aforementioned. This principle is formalised by
the Markov condition \parencite[see for example][]{Lauritzen1996}.
This assumption is considered to be %
mild. Any distribution induced by an acyclic SCM
satisfies the Markov condition with respect to the corresponding graph \parencite{Lauritzen1990, Pearl2009}.
The second assumption (often referred to
as faithfulness),
states that
any
(conditional) independence
between random variables is implied by the graph structure
\parencite{Spirtes2000}.
For example,
if two variables are independent, then neither does cause the other nor do they share a common cause.
Both assumptions together
establish a one-to-one correspondence
between conditional independences in the distribution
and graphical separation properties between the corresponding nodes.

The back-bone of the constraint-based causal discovery
algorithms such as the PC algorithm
is to test for marginal and conditional (in)dependences in observed data and to
find all graphs that encode the same list of separation statements~\parencite{Spirtes2000,Pearl2009}.
This allows us to
infer a so-called Markov equivalence class of graphs: all of its members encode the same set
of conditional independences.
It has been shown that
two directed acyclic graphs
(assuming that all nodes are observed)
are Markov equivalent if and only if they have the same skeleton and v-structures $\to{\circ}\gets$~\parencite{verma1990equivalence}.
Allowing for hidden variables, as done by the FCI algorithm, for example, enlarges the class of equivalent graphs and the output is usually less informative \parencite{Spirtes2000}.

The following example further illustrates the idea of a constraint-based search.
For simplicity, we assume a linear Gaussian setting, so that (conditional) independence coincides with vanishing (partial) correlation.
Say we observe $X$, $Y$, and $Z$.
Assume that the partial correlation between $X$ and $Z$ given $Y$ vanishes while
none of the other correlations and partial correlations vanish.
Under the Markov and faithfulness assumptions there are multiple causal structures that are compatible with those constraints, such as $X \to Y \to Z$, $X \gets Y \gets Z$, $X \gets Y \to Z$, or
\begin{center}
\begin{tikzpicture}[baseline={([yshift=.8ex]current bounding box.south)}]
    \node (h) at(1,1) {$H$};
    \node (x) at(0,0) {$X$};
    \node (y) at(2,0) {$Y$};
    \node (z) at(4,0) {$Z$};

    \draw[->] (h) -- (y);
    \draw[->] (h) -- (x);
    \draw[->] (y) -- (z);
\end{tikzpicture},
$\quad$
or
$\quad$
\begin{tikzpicture}[baseline={([yshift=.8ex]current bounding box.south)}]
    \node (h) at(1,1) {$H$};
    \node (x) at(0,0) {$X$};
    \node (y) at(2,0) {$Y$};
    \node (z) at(4,0) {$Z$};

    \draw[->] (h) -- (y);
    \draw[->] (h) -- (x);
    \draw[->] (x) -- (y);
    \draw[->] (y) -- (z);
\end{tikzpicture},
\end{center}
where $H$ is unobserved.
Still, the correlation pattern rules out certain other causal structures.
For example,
neither $X \to Y \gets Z$
nor $X \gets H \to Y \gets Z$
can be the correct graph structure since either case would imply that
$X$ and $Z$ are uncorrelated (and $X \independent Z \mid Y$ is not satisfied).

Variants of the above setting were considered in neuroimaging where a randomised experimental stimulus or time-ordering was used to further disambiguate between the remaining possible structures~\parencite{grosse2016identification,weichwald2016recovery,weichwald2016merlin,mastakouri2019selecting}.
Constraint-based causal inference methodology also clarifies the interpretation of encoding and decoding analyses in neuroimaging and has informed a refined understanding of the neural dynamics of probabilistic reward prediction and an improved functional atlas~\parencite{weichwald2015causal,bach2017whole,varoquaux2018atlases}.

Direct applications of this approach in cognitive neuroscience are difficult, not only due to the key challenges discussed in Section~\ref{sec:challenges}, but also due to indirect and spatially smeared neuroimaging measurements that effectively spoil conditional independences.
In the linear setting, there are recent advances that explicitly tackle the problem of inferring the causal structure between latent variables, say the neuronal entities, based on observations of recorded variables~\parencite{silva2006learning}.
Further practical challenges include
the
difficulty of testing for non-parametric conditional independence~\parencite{Shah2020}
and near-faithfulness violations \parencite{Uhler2013}.

\paragraph{Score-based methods.}
Instead of directly exploiting the (conditional) independences to inform our inference about the causal graph structure, score-based methods assess different graph structures by their ability to fit observed data~\parencite[see for example][]{Chickering2002}.
This approach is motivated by the idea that graph structures that encode the wrong (conditional) independences will also result in bad model fit.
Assuming a parametric model class, we can evaluate the log-likelihood of the data
and score different candidate graph structures by the Bayesian Information Criterion, for example.
The number of possible graph structures to search over
grows super-exponentially.
That combinatorial difficulty
can be
dealt with by applying greedy search procedures
that usually, however, do not come with finite sample guarantees.
Alternatively, \textcite{zheng2020learning} exploit
an algebraic characterisation of graph structures
to maximise a score over acyclic graphs
by solving a continuous optimisation problem.
The score-based approach relies on correctly specifying the model class.
Furthermore, in the presence of hidden variables,
the search space grows even larger and model scoring is complicated by the need to marginalise over those hidden variables~\parencite{jabbari2017discovery}.

\paragraph{Restricted structural causal models.}
Another possibility is to restrict the class of functions in the structural assignments and the noise distributions.
Linear non-Gaussian acyclic models~\parencite{shimizu2006a}, for example,
assume that the structural assignments are linear
and the noise distributions are non-Gaussian.
As for independent component analysis, identifiability
of the causal graph follows from the Darmois-Skitovich
theorem~\parencite{darmois, skitovich}.
Similar results hold for nonlinear models with additive noise~\parencite{hoyer2008nonlinear, Zhang2009, Peters2014jmlr,Buehlmann2014annals}
or linear Gaussian models when the
error variances of the different variables
are assumed to be equal~\parencite{Peters2014biometrika}.
The additive noise assumption is a powerful, yet restrictive, assumption that
may be violated in practical applications.

\paragraph{Dynamic causal modelling (DCM).}
We may have prior beliefs about the existence and direction of some of the edges.
Incorporating these
by careful specification of the priors is an explicit modelling step in DCM~\parencite{valdes2011effective}.
Given such a prior, we may prefer one model over the other among the two observationally equivalent models presented in Section~\ref{sec:modelequivalence}, for example.
Since the method's outcome relies on this prior information, any disagreement
on the validity of that prior information necessarily yields a
discourse about
the method's outcome~\parencite{lohmann2012critical}.
Further, a simulation study raised concerns regarding the validity of the model selection procedure in DCM~\parencite{friston2003dynamic,lohmann2012critical,friston2013model,breakspear2013dynamic,lohmann2013response}.

\paragraph{Granger causality.}
Granger causality is among the most popular approaches for the analysis of connectivity between time-evolving processes.
It
exploits the existence of
time and the fact that causes precede their effects.
Together with its non-linear extensions it has been considered for the analysis of neuroimaging data with applications to
electro-encephalography (EEG)
and fMRI data~\parencite{marinazzo2008kernel,marinazzo2011nonlinear,stramaglia2012expanding,stramaglia2014synergy}.
The idea is sometimes wrongly described as follows:
If including the past of $Y_t$ improves our prediction of $X_t$ compared to a prediction that is only based on the past of $X_t$ alone, then $Y$ Granger-causes $X$.
\textcite{granger1969investigating} himself put forward a more careful definition that includes a reference to all the information in the universe:
If the prediction of $X_t$ based on all the information in the universe up to time $t$ is better than
the prediction where we use all the information in the universe up to time $t$ apart from the past of $Y_t$, then $Y$ Granger-causes $X$.
In practice, we
may instead resort to a multivariate formulation of Granger causality.
If all
relevant variables are observed (often referred to as causal sufficiency),
there is a close correspondence between Granger causality and the constraint-based approach~\parencite[Chapter 10.3.3]{peters2017elements}.
Observing all relevant variables, however, is a strong assumption which is most likely violated for data sets in cognitive neuroscience.
While Granger causality may be combined with a goodness-of-fit test
to at least partially detect the existence of confounders \parencite{Peters2013nips},
it is commonly applied as a computationally efficient black box approach that
always outputs a result.
In the presence of instantaneous effects (for example, due to undersampling) or hidden variables, these results may be erroneous~\parencite[see, for example,][]{sanchez2019estimating}.

\paragraph{Inferring causes of a target variable.}
We now consider a problem that is arguably simpler than
inferring the full causal graph:
identifying the causes of some target variable of interest.
As outlined in the running examples in Section~\ref{sec:runningexample}, we assume that we have observations of the variables $T, Y, X_1, \ldots, X_d$, where $Y$ denotes the target variable.
Assume that there is an unknown structural causal model that includes the variables $T, Y, X_1, \ldots, X_d$ and that describes the data generating process well.
To identify the variables among $X_1, \ldots, X_d$ that cause $Y$,
it does not suffice to
regress $Y$ on $X_1, \ldots, X_d$.
The following example  of an SCM shows that
a good predictive model for $Y$ is not necessarily
a good interventional model for $Y$. Consider

\begin{minipage}[b]{.5\textwidth}
\begin{align*}
X_1 &:= N_1 \\
Y &:= X_1 + N_Y\\
X_2 &:= 10\cdot Y + N_2
\end{align*}
\end{minipage}%
\begin{minipage}[b]{.5\textwidth}
\begin{tikzpicture}
\node (x1) at(1,.5) {$X_1$};
\node (x2) at(3,-.5) {$X_2$};
\node (y) at(2,0) {$Y$};
\draw[->] (x1) -- (y);
\draw[->] (y) -- (x2);
\end{tikzpicture}
\end{minipage}
where $N_1,N_2,N_Y$ are mutually independent standard-normal noise variables.
$X_2$ is a good predictor for $Y$, but $X_2$ does not have any causal influence on $Y$: the distribution of $Y$ is unchanged upon interventions on $X_2$.

Recently, causal discovery methods
have been proposed that
aim to infer the causal parents of $Y$
if we are given data from different environments, that is, from different experimental conditions, repetitions, or different subjects.
These methods exploit a distributional robustness property of causal models and are described in
Section~\ref{sec:robustness}.

\paragraph{Cognitive function versus brain activity as the target variable.}
When we are interested in inferring  direct causes of a target variable
$Y$, it can be useful to include background knowledge.
Consider our Running Example~A (cf.\ Section~\ref{sec:runningexample} and Figure~\ref{fig:toyexampleA}) with reaction time as the target variable
and assume we are interested in inferring which of the variables measuring neuronal activity are causal for the reaction time $Y$.
We have argued in the preceding paragraph that if a variable $X_j$ is predictive of $Y$, it does not necessarily have to be causal for $Y$.
Assuming, however, that we can exclude that the cognitive function `reaction time' causes brain activity (for example, because of time ordering), we obtain the following simplification:
every $X_j$ that is predictive of $Y$, must be an indirect or direct cause of $Y$, confounded with $Y$, or a combination of both.
This is different if our target variable is a neuronal entity as in Running Example~B (cf.~Figure~\ref{fig:toyexampleB}).
Here, predictive variables can be either ancestors of $Y$, confounded with $Y$, descendants of $Y$, or some combination of the aforementioned (these statements follow from
the Markov condition).

\section{Two challenges for causality in cognitive neuroscience}
\label{sec:challenges}
Performing
causal inference on measurements of neuronal activity comes with several challenges, many of which have been discussed in the literature (cf.\ Section~\ref{sec:existingwork}).
In the following two subsections we explicate two challenges that we think deserve special attention.
In Section~\ref{sec:robustness}, we elaborate on how distributional robustness across environments, such as different recording sessions or subjects, can serve as a guiding principle for tackling those challenges.

\subsection{Challenge 1: The scarcity of targeted interventional data}\label{sec:challenge1}

In Section~\ref{sec:modelequivalence} we discussed that
different causal models may induce the same observational distribution while they make different predictions about the effects of interventions.
That is, observationally equivalent models need not be interventionally equivalent.
This implies that
some models can only be refuted when we observe the system under interventions
which perturb some specific variables in our model.
In contrast to broad perturbations of the system,
we call targeted interventions those
for which the intervention target is known and
for which we can list the intervened-upon variables in our model,
say ``$X_1, X_3, X_8$ have been intervened upon.''
Even if some targeted interventions are available, there may still be multiple models that are compatible with all observations obtained under those available interventions.
In the worst case,
a sequence of up to $d$
targeted
interventional experiments
may be required to distinguish between the possible causal structures
over $d$ observables $X_1,\dots,X_d$
when the existence of unobserved variables cannot be excluded
while assuming Markovianity, faithfulness, and acyclicity~\parencite{eberhardt2013experimental}.
In general, the more interventional scenarios are available to us, the more causal models we can falsify and the further we can narrow down the set of causal models compatible with the data.

Therefore, the scarcity of targeted interventional data
is a barrier to causal inference in cognitive neuroscience.
Our ability to intervene on neural entities such as the BOLD level or oscillatory bandpower in a brain region is limited and so is our ability
to either
identify the right causal model from interventional data
or to test causal hypotheses that are made in the literature.
One promising avenue are non-invasive brain stimulation techniques such as
transcranial magnetic or direct/alternating current stimulation
which modulate neural activity by creating a field inside the brain~\parencite{nitsche2008transcranial,herrmann2013transcranial,bestmann2017transcranial,kar2020transcranial}.
Since the stimulation acts broadly and its neurophysiological effects are not yet fully understood, transcranial stimulation cannot be understood as targeted intervention on some specific neuronal entity in our causal model~\parencite{antal2016transcranial,vosskuhl2018non}.
The inter-individual variability in response to stimulation further impedes its direct use for probing causal pathways between brain regions~\parencite{lopez2014inter}.
\textcite{bergmannFCinferring} review the obstacles to inferring causality from non-invasive brain stimulation studies
and provide guidelines to attenuate the aforementioned.
Invasive stimulation techniques,
such as deep brain stimulation relying on electrode implants~\parencite{mayberg2005deep},
may enable temporally and spatially more fine-grained perturbations
of neural entities.
\textcite{dubois2017causal} exemplify how
to revise causal structures inferred form observational neuroimaging data on a larger cohort
through direct stimulation of specific brain regions and concurrent fMRI on a smaller cohort of neurosurgical epilepsy patients.
In non-human primates, concurrent optogenetic stimulation with whole-brain fMRI
had been used to map the wiring of the medial prefrontal cortex~\parencite{liang2015mapping,lee2010global}.
Yet, there are ethical barriers to large-scale invasive brain stimulation studies
and it may not be exactly clear how an invasive stimulation corresponds to an intervention on, say, the BOLD response measured in some voxels.
We thus believe that targeted interventional data will remain a scarcity
due to physical and ethical limits to non-invasive and invasive brain stimulation.

Consider the following variant of our Running Example B (cf.\ Section~\ref{sec:runningexample}).
Assume that
(a)~the consumption of alcoholic beer $T$ slows neuronal activity in the brain regions $X_1$, $X_2$, and $Y$,
(b)~$X_2$ is a cause of $X_1$,
and
(c)~$X_2$ is a cause of $Y$.
Here, (a)~could have been established by randomising $T$, whereas (b)~and (c)~may be background knowledge.
Nothing is known, however, about the causal relationship between $X_1$ and $Y$ (apart from the confounding effect of $X_2$).
The following graph summarises these causal relationships between the variables:
\begin{center}
\resizebox{.5\textwidth}{!}{
\begin{tikzpicture}
    \node (X) at(0,0) {{\reflectbox{\includegraphics[keepaspectratio,width=7cm]{brain.png}}}};

    \node[node] (T) at(-6,0) {$T$};
    \node[nodeh,draw=gray,text=gray] (H) at(2.5,3) {$H$};
    \node[node] (x1) at(-2,-1.5) {$X_1$};
    \node[node] (x2) at(0,1.5) {$X_2$};
    \node[node] (x3) at(2,0) {$Y$};

    \draw[->,la] (T) -- (x1);
    \draw[->,la] (T) -- (x2);
    \draw[->,la] (T) -- (x3);
    \draw[->,la,gray] (H) -- (x1);
    \draw[->,la,gray] (H) -- (x2);
    \draw[->,la,gray] (H) -- (x3);
    \draw[->,la] (x2) -- (x3);
    \draw[->,la,red] (x3) -- (x1);
    \draw[->,la] (x2) -- (x1);
    \draw[->,la,red] (x1) -- (x3);
    \node[fill=white,circle,text=red] (q2) at(0,-.75) {{\LARGE ?}};
\end{tikzpicture}
}
\end{center}
Assume we establish on observational data that there is a dependence between $X_1$ and $Y$ and that we cannot render these variables conditionally independent by conditioning on any combination of the remaining observable variables $T$ and $X_2$.
Employing the widely accepted Markov condition,
we can conclude that either
$X_1\to Y$,
$X_1\gets Y$,
$X_1\gets H \to Y$ for some unobserved variable $H$,
or some combination of the aforementioned settings.
Without any further assumptions, however,
these models
are observationally equivalent.
That is, we cannot refute any of the above possibilities
based on observational data alone.
Even randomising $T$ does not help:
The above models are interventionally equivalent with respect to
interventions on~$T$.
We could %
apply one of the
causal discovery methods described in Section~\ref{sec:discovery}.
All of these methods, however, employ further assumptions on the
data generating process that go beyond the Markov condition.
We may deem some of those assumptions implausible given prior knowledge about the system.
Yet, in the absence of targeted interventions on $X_1$, $X_2$ or $Y$, we can neither falsify candidate models obtained by such methods nor can we test all of the underlying assumptions.
In Section~\ref{sec:interventional},
we illustrate how we may benefit from heterogeneity in the data, that is,
from interventional data where the intervention target is unknown.

\subsection{Challenge 2: Finding the right variables}\label{sec:challenge2}
Causal discovery often starts by considering observations of some variables $Z_1,\dots,Z_d$ among which we wish to infer cause-effect relationships, thereby implicitly assuming that those variables are defined or constructed in a way that they can meaningfully be interpreted as causal entities in our model.
This, however, is not necessarily the case in neuroscience.
Without knowing how higher-level causal concepts emerge from lower levels, for example, it is hard to imagine how to make sense and use of a causal model of the $86$ billion neurons in a human brain~\parencite{Herculano-Houzel10661}.
One may hypothesise that
a model of averaged neuronal activity in
distinct functional brain regions may be pragmatically useful to reason about the effect of different treatments and to understand the brain.
For such an approach we need to find the right transformation of the high-dimensional observed variables to obtain the right variables for a causal explanation of the system.%

The problem of relating %
causal models with different granularity and
finding the right choice of variable transformations that
enable
causal reasoning %
has received attention in the causality literature also outside of neuroscience applications.
\textcite{eberhardt2016green} fleshes out an instructive two-variable example that demonstrates that the choice of variables for causal modelling may be underdetermined even if interventions were available.
For a wrong choice of variables our ability to causally reason about a system breaks.
An example of this is the historic
debate
about whether a high cholesterol diet was beneficial or harmful with respect to heart disease.
It can be partially explained by
an ambiguity of how exactly
total cholesterol is manipulated.
Today, we know that
low-density lipoproteins %
and high-density lipoproteins have opposing effects on heart disease risk.
Merging these variables together to total cholesterol does not yield a variable with a well-defined intervention:
Referring to an
intervention on total cholesterol does
not specify what part of the intervention
is due to a change in
low-density lipoproteins
(LDL)
versus high-density lipoproteins
(HDL).
As such,
only including
total cholesterol instead of LDL and HDL
may therefore be regarded as a too coarse-grained variable representation that breaks a model's causal semantics, that is, the ability to map every intervention to a well-defined
interventional
 distribution~\parencite{spirtes2004causal,steinberg2007the,truswell2010cholesterol}.

Yet, we may sometimes prefer to transform micro variables into macro variables.
This can result in a concise summary of the causal information that abstracts away detail, is easier to communicate and operationalise, and more effectively represents the information necessary for a certain task~\parencite{hoel2013quantifying,hoel2017when,weichwald2019pragmatism}; for example, a causal model over $86$ billion neurons may be unwieldy for a brain surgeon aiming to identify and remove malignant brain tissue guided by the cognitive impairments observed in a patient.
\textcite{rubenstein2017causal} formalise a notion of exact transformations that ensures causally consistent reasoning between two causal models where the variables in one model are transformations of the variables in the other.
Roughly speaking, two models are considered causally consistent
if the following two ways to reason about how the distribution of the macro-variables changes upon a macro-level intervention agree with one another:
(a) find an intervention on the micro-variables that corresponds to the considered macro-level intervention,
and consider the macro-level distribution implied by the micro-level intervention, and
(b) obtain the interventional distribution directly within the macro-level structural causal model sidestepping any need to refer to the micro-level.
If the two resulting distributions agree with one another for all (compositions of) interventions,
then the two models are said to be causally consistent and we can view the macro-level as an exact transformation of the micro-level causal model that preserves its causal semantics.
A formal exposition of the framework and its technical subtleties can be found in
the aforementioned work.
Here, we revisit a variant of the cholesterol example for an illustration
of what it entails for two causal models to be causally consistent
and illustrate a failure mode:
Consider variables $L$ (LDL), $H$ (HDL), and $D$ (disease),
where
$D := H - L + N_D$
for $L, H, N_d$ mutually independent random variables.
Then a model based on the transformed variables $T = L + H$ and $D\equiv D$ is in general not causally consistent with the original model:
For $(l_1,h_1) \neq (l_2,h_2)$
with $l_1+h_1 = l_2+h_2$
the interventional distributions induced by the micro-level model corresponding to setting
$L:=l_1$ and $H:=h_1$ or alternatively $L:=l_2$ and $H:=h_2$
do in general not coincide
due to the differing effects of $L$ and $H$ on $D$.
Both interventions correspond to the same level of $T$ and the intervention setting
$T := t$
with
$t = l_1 + h_1 = l_2 + h_2$
in the macro-level model.
Thus, the distributions obtained from reasoning
(a) and (b) above do not coincide.
If, on the other hand, we had
$\widetilde{D} := H + L + N_D$,
then we could
indeed use a macro-level model where we consider
$T=H+L$ to reason about
the distribution of $\widetilde{D}$ under the intervention $\doop{T:= t}$
without running into conflict with the interventional distributions implied by all corresponding interventions in the micro-level model.
This example can analogously be considered in the context of our running examples (cf.\ Section~\ref{sec:runningexample}):
Instead of LDL, HDL, and disease one could alternatively
think of
some neuronal activity
($L$)
that delays motor response,
some neuronal activity
($H$)
that increases attention levels,
and the detected reaction time
($D$)
assessed by subjects performing a button press;
the scenario then translates into how causal reasoning about the cause of
slowed reaction times is hampered once we give up
on considering $H$ and $L$ as two separate neural entities
and instead try to reason about the average activity $T$.
\textcite{janzing2018structural} observe similar problems for causal reasoning when aggregating variables and show that the observational and interventional stationary distributions of a bivariate autoregressive processes cannot in general be described by a two-variable causal model.
A recent line of research focuses on developing a notion of approximate transformations of causal models~\parencite{beckers2019abstracting,beckers2019approximate}.
While there exist first approaches to learn discrete causal macro-variables from data~\parencite{chalupka2015visual,chalupka2016multi}, we are unaware of any method that is generally applicable and learns causal variables from complex high-dimensional data.

In cognitive neuroscience, we commonly treat large-scale brain networks or brain systems as
causal entities and then proceed to infer interactions between those~\parencite{yeo2011the,power2011functional}.
\textcite{smith2011network} demonstrate that
this should be done with caution:
Network identification is strongly susceptible to slightly wrong or different definitions of the regions of interest (ROIs) or the so-called atlas.
Analyses based on Granger causality depend on the level of spatial aggregation and were shown to reflect the intra-areal properties instead of the interactions among brain regions if an ill-suited aggregation level is considered~\parencite{chicharro2014algorithms}.
Currently,
there does not seem to be
consensus as to which macroscopic entities and brain networks are the right ones to (causally) reason about cognitive processes~\parencite{uddin2019towards}.
Furthermore,
the observed variables themselves are already aggregates:
A single fMRI voxel or the local field potential at some cortical location reflects the activity of thousands of neurons \parencite{logothetis2008we,einevoll2013modelling};
EEG recordings are commonly considered a linear superposition of cortical electromagnetic activity which has spurred the development of blind source separation algorithms that try to invert this linear transformation to recover the underlying cortical variables~\parencite{nunez2006electric}.

\section{Causality and leveraging robustness}
\label{sec:robustness}

\subsection{Robustness of causal models}\label{sec:robustcausalmodels}
The concept of causality is
linked to
invariant models and
distributional robustness.
Consider again the setting with
a target variable $Y$ and
covariates $X_1, \ldots, X_d$,
as described in the running examples in Section~\ref{sec:runningexample}. %
Suppose %
that the system is observed in different environments.
Suppose
further that the generating process can be described by an SCM,
that
$\pa{Y} \subseteq \{X_1, \ldots, X_d\}$
are the causal parents of $Y$, and
that
the different
environments
correspond to
different
interventions on some of the covariates,
while we neither (need to) know the
interventions' targets nor its precise form.
In our reaction time example,
the two environments may
represent two subjects
(say, a left-handed subject right after having dinner and  a trained race car driver just before a race)
that differ in
the mechanisms for
$X_1$, $X_3$, and $X_7$.
Then the joint distribution over
$Y, X_1, \ldots, X_d$ may be different between the environments and also the marginal distributions may vary.
Yet, if the interventions do not act directly on $Y$
the causal
model is invariant in the following sense:
the conditional distribution of
$Y\,|\, \mathbf{X}_{\pa{Y}}$
is the same in all environments.
In the reaction time examples this could translate to the neuronal causes
that facilitate fast (versus slow)
reaction times to be the same across subjects.
This invariance can be formulated in different ways.
For example, we have for all $k$ and $\ell$,
where $k$ and $\ell$ denote the indices of two environments, and for
almost all $x$
\begin{equation}\label{eq:invariance}
(Y^k \,|\, \mathbf{X}^k_{\pa{Y}} = x)
=
(Y^\ell \,|\, \mathbf{X}^\ell_{\pa{Y}} = x)
\qquad \text{in distribution}.
\end{equation}
Equivalently,
\begin{equation}\label{eq:invariance2}
E \independent Y \,|\, \mathbf{X}_{\pa{Y}},
\end{equation}
where the variable $E$ represents the environment.
In practice,
we often work with model classes such
as linear or logistic regression
for modelling the conditional distribution
$Y \,|\, \mathbf{X}_{\pa{Y}}$.
For such model classes, the above statements
simplify.
In case of linear models, for example, Equations~\eqref{eq:invariance}
and~\eqref{eq:invariance2}
translate to regression coefficients and error variances being equal across different environments.

For an example,
consider a system
that, for environment $E=1$,
is governed by the following structural assignments

\begin{minipage}{.3\textwidth}
SCM for $E=1$:
\begin{align*}
X_1 &:= N_1 \\
X_2 &:= 1\cdot X_1 + N_2 \\
X_3 &:= N_3 \\
Y &:= X_1 + X_2 + X_3 + N_Y\\
X_4 &:= Y + N_2
\end{align*}
\end{minipage}%
\begin{minipage}{.2\textwidth}
\hspace{.5em}
\begin{tikzpicture}
\node (x1) at(0,1) {$X_1$};
\node (x2) at(1,-1) {$X_4$};
\node (x3) at(2,1) {$X_3$};
\node (x4) at(1,1) {$X_2$};
\node (y) at(1,0) {$Y$};
\draw[->] (x1) -- (y);
\draw[->] (x1) -- (x4);
\draw[->] (x4) -- (y);
\draw[->] (y) -- (x2);
\draw[->] (x3) -- (y);
\end{tikzpicture}
\end{minipage}%
\begin{minipage}{.5\textwidth}
\includegraphics[keepaspectratio,width=\textwidth]{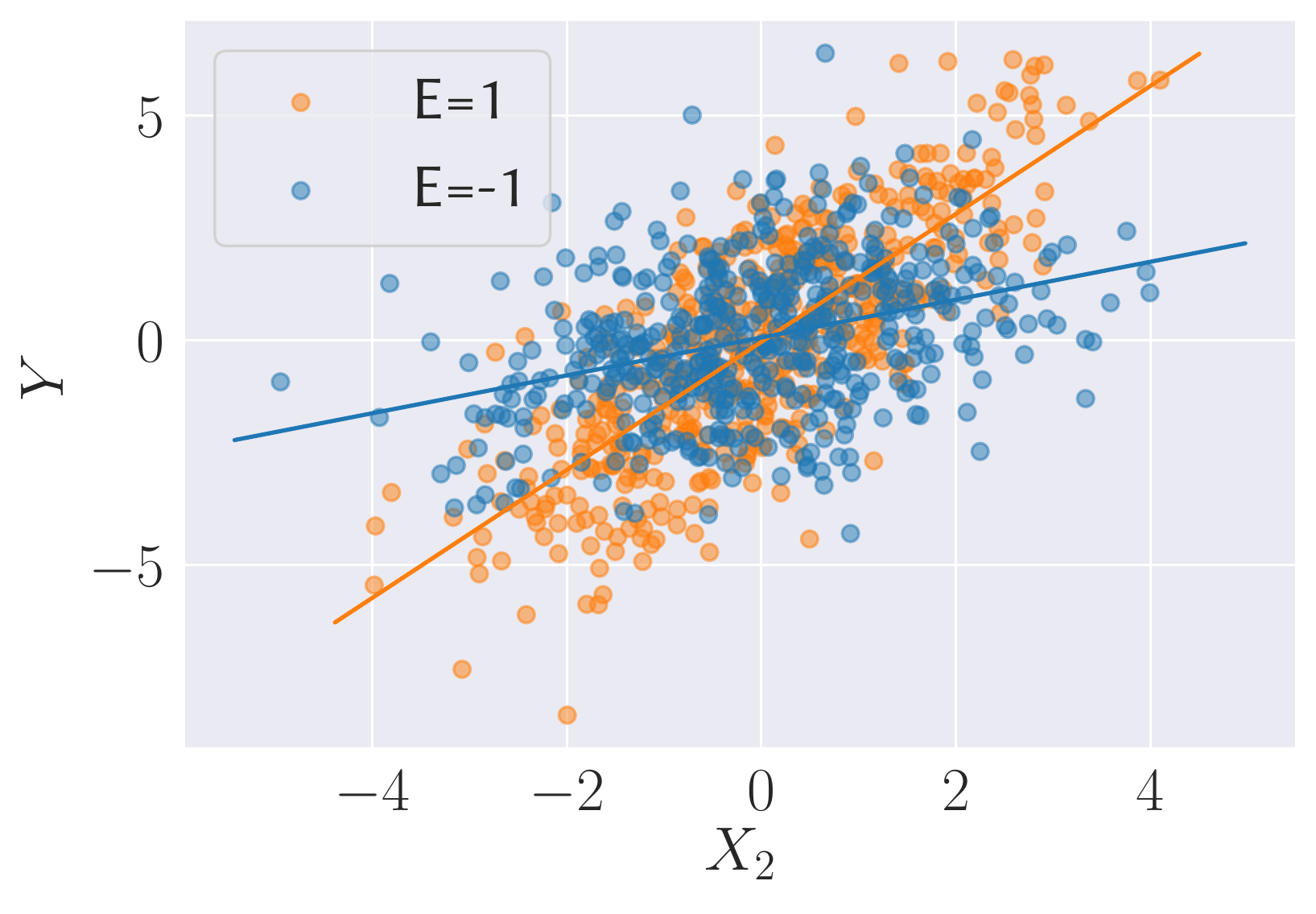}
\end{minipage}
\-\\
with $N_1, N_2, N_3, N_4, N_Y$ mutually independent and standard-normal,
and where environment $E=-1$ corresponds to an intervention
changing the weight of $X_1$ in the assignment for $X_2$ to $-1$.
Here, for example, $\{X_1, X_2, X_3\}$ and $\{X_1, X_2, X_4\}$
are
so-called
invariant sets:
the conditionals $Y|X_1,X_2,X_3$ and $Y|X_1,X_2,X_4$ are the same in both environments.
The invariant models $Y|X_1,X_2,X_3$ and $Y|X_1,X_2,X_4$ generalise to a new environment $E=-2$,
which changes the same weight to $-2$, in that they would still predict well.
Note that $Y|X_1,X_2,X_4$ is a non-causal model.
The lack of invariance of $Y|X_2$ is illustrated by the different regression lines in the scatter plot on the right.

The validity of~\eqref{eq:invariance} and~\eqref{eq:invariance2}
follows from the fact that the interventions
do not act on $Y$ directly
and can
be proved using the
equivalence of Markov conditions \parencite[][Section 6.6]{Lauritzen1996,Peters2016jrssb}.
Here, we try to argue that it also makes sense intuitively.
Suppose that someone proposes to have found
a complete
causal model for a target variable~$Y$,
using certain covariates $\mathbf{X}_{S}$ (for $Y$, we may
again
think of the reaction time in Example~A).
Suppose that fitting that model for different subjects yields
significantly different model fits
--
maybe even with different signs
for the causal effects from variables in $\mathbf{X}_S$ to $Y$
such that $E \independent Y \mid \mathbf{X}_{S}$ is violated.
In this case,
we would
become
sceptical about whether the proposed model is indeed
a complete causal model.
Instead, we might suspect that the model is missing an important
variable describing how reaction time
depends on brain activity.

In practice,
environments can
represent different
sources of heterogeneity.
In a cognitive neuroscience setting,
environments may be thought of as different subjects
who react differently, yet not arbitrarily so (cf.\ Section~\ref{sec:whycausality}),
to varying levels of alcohol consumption.
Likewise, different experiments that are thought to involve the same
cognitive processes may be thought of as environments;
for example,
the relationship `neuronal activity $\to$ reaction time' (cf.\ Example A, Section~\ref{sec:runningexample})
may be expected to translate from
an experiment that compares reaction times
after consumption of alcoholic versus non-alcoholic beers
to another experiment where subjects are exposed to Burgundy wine versus grape juice.
The key assumption is that
the environments do not alter the
mechanism of $Y$%
—that is, $f_Y(\pa{Y}, N_Y)$—%
directly or, more formally,
there are no interventions on $Y$.
To test
whether a set of covariates is invariant, as described in~\eqref{eq:invariance} and~\eqref{eq:invariance2},
no causal
background knowledge is required.

The above invariance
principle is also known as
`modularity' or
`autonomy'.
It has been discussed
not only in the field of
econometrics
\parencite{Haavelmo1944, Aldrich1989, Hoover2006},
but also in philosophy of science.
\textcite{woodward2005making}
discusses how the  invariance idea rejects
that `either a generalisation is a law or else is
purely accidental'.
In our notion, the criteria~\eqref{eq:invariance}
and~\eqref{eq:invariance2} depend on the environments $E$. In particular,
a model may be invariant with respect to some changes, but not with respect to others.
In this sense, robustness and invariance  should
always be thought with respect to
a certain set of changes.
\textcite{woodward2005making}
introduces the possibility
to talk about various degrees of invariance,
beyond the mere existence or absence of invariance, while
acknowledging that mechanisms that
are sensitive even to mild changes in the background conditions are usually considered as not scientifically interesting.
\textcite{Cartwright2003}
analyses the relationship between
invariant and causal relations
using linear deterministic systems and draws conclusions analogous to the ones discussed above.
In the context of the famous Lucas critique \parencite{Lucas},
it is debated to which extent
invariance can be used for predicting the effect of
changes in economic policy
\parencite{Cartwright2009}:
Economy consists of many individual
players who are capable of adapting their behaviour to a change in policy.
In cognitive neuroscience, we believe that the situation is different.
Cognitive mechanisms
do change and adapt, but not necessarily arbitrarily quickly.
Some cognitive mechanism of an individual at the same day can be assumed to be invariant with respect to changes in the visual input, say.
Depending on the precise setup, however,
we may expect moderate changes of the mechanisms,
say, for example, the development of cognitive function in children or learning effects.
In other settings, where mechanisms may be subject to
arbitrary large changes, scientific insight seems impossible (see Section~\ref{sec:whycausality}).

Recently,
the principle of invariance has
also received
increasing attention in the statistics
and machine learning community
\parencite{Schoelkopf2012icml,Peters2016jrssb,Arjovsky2019}.
It
can also be applied to models
that do not have the form of an SCM.
Examples include
dynamical models
that are governed by differential equations
\parencite{Pfister2019pnas}.

\subsection{Distributional robustness and scarcity of interventional data}\label{sec:interventional}

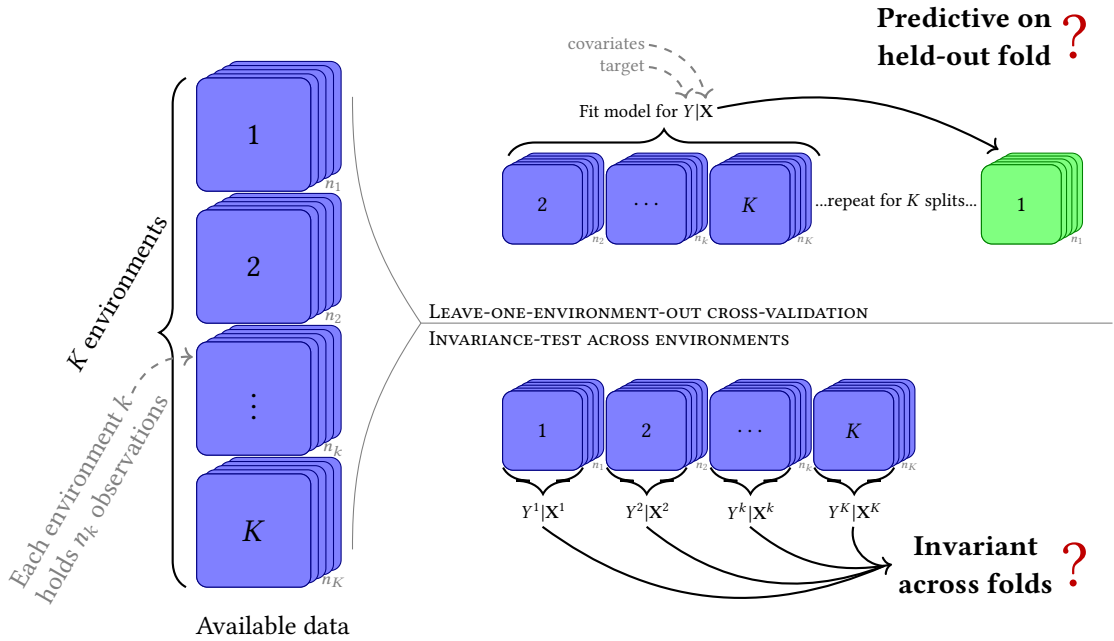
\begin{figure}
\centering
\begin{tikzpicture}
\begin{scope}[shift={(-.8,0)}]
\drawdatabox{0}{0}{black!50!blue}{blue!50!white}{$1$}{$n_1$}

\drawdatabox{0}{-1.75}{black!50!blue}{blue!50!white}{$2$}{$n_2$}
\drawdatabox{0}{-3.5}{black!50!blue}{blue!50!white}{$\vdots$}{$n_k$}
\drawdatabox{0}{-5.25}{black!50!blue}{blue!50!white}{$K$}{$n_K$}

\node at (0.25,-6.5) {\footnotesize Available data};

\draw [decorate,decoration={brace,amplitude=10pt},xshift=-4pt,yshift=0pt,thick]
(-.75,-6) -- (-.75,.75) node [black,rotate=60,midway,yshift=2.5em,align=left]
{{\footnotesize $K$ environments}};

\draw[->,gray,dashed,thick] (-1.6,-3.43) to[bend right=-35] (-.8, -2.95);
\node[gray,align=left,rotate=60] at (-2.15, -4.73) {
{\footnotesize Each environment $k$}\\[-.25em]
{\footnotesize holds $n_k$ observations}};
\end{scope}

\begin{scope}[shift={(0,0.3)}]
\begin{scope}[shift={(3,0)},scale=0.7,every node/.style={scale=0.7}]
\drawdatabox{0}{-1.75}{black!50!blue}{blue!50!white}{$2$}{$n_2$}
\drawdatabox{1.95}{-1.75}{black!50!blue}{blue!50!white}{$\hdots$}{$n_k$}
\drawdatabox{3.9}{-1.75}{black!50!blue}{blue!50!white}{$K$}{$n_K$}

\draw [decorate,decoration={brace,amplitude=10pt},xshift=-4pt,yshift=0pt,thick]
(-.5,-.83) -- (5.35,-.83);

\node[anchor=center,align=left] at (1.95,0) {\footnotesize Fit model for $Y|\mathbf{X}$};

\draw[->,gray,dashed,thick] (2,.85) to[bend left=35] (2.7, .25);
\node[gray] at (1.5, .83) {\footnotesize target};

\draw[->,gray,dashed,thick] (2, 1.27) to[bend left=35] (3.07, .25);
\node[gray,align=center] at (1.2, 1.3) {\footnotesize covariates};

\node at (6.65,-1.7) {\footnotesize ...repeat for $K$ splits...};

\drawdatabox{9}{-1.75}{black!50!green}{green!50!white}{$1$}{$n_1$}
\end{scope}

\node[align=center] at (8.55,1.01) {\small\bfseries Predictive on\\ \small\bfseries held-out fold};
\node[red!75!black] at(10,1) {\Huge ?};
\draw[->,thick] (5.3,0.05) to[bend left=20] (9, -.5);
\end{scope}

\node[anchor=center] at(4.4,-2.35) {\scriptsize \textsc{Leave-one-environment-out cross-validation}};

\draw[-,gray] (1.4,-2.5) to[bend right=-17] (0.5,.5) -- (0.5,.5);
\draw[-,gray] (1.4,-2.5) to[bend left=-17] (0.5,-5.5) -- (0.5,-5.5);
\draw[-,gray] (1.4,-2.5) -- (10.5,-2.5);

\node[anchor=center] at(3.88,-2.7) {\scriptsize \textsc{Invariance-test across environments}};

\begin{scope}[shift={(0,-.2)}]
\begin{scope}[shift={(3,-2.5)},scale=0.7,every node/.style={scale=0.7}]
\drawdatabox{0}{-1.75}{black!50!blue}{blue!50!white}{$1$}{$n_1$}
\drawdatabox{1.95}{-1.75}{black!50!blue}{blue!50!white}{$2$}{$n_2$}
\drawdatabox{3.9}{-1.75}{black!50!blue}{blue!50!white}{$\hdots$}{$n_k$}
\drawdatabox{5.85}{-1.75}{black!50!blue}{blue!50!white}{$K$}{$n_K$}
\draw [decorate,decoration={brace,amplitude=10pt},xshift=-4pt,yshift=0pt,thick]
(.9,-2.45) -- (-.6,-2.45);
\draw [decorate,decoration={brace,amplitude=10pt},xshift=-4pt,yshift=0pt,thick]
(.9+1.95,-2.45) -- (-.6+1.95,-2.45);
\draw [decorate,decoration={brace,amplitude=10pt},xshift=-4pt,yshift=0pt,thick]
(.9+2*1.95,-2.45) -- (-.6+2*1.95,-2.45);
\draw [decorate,decoration={brace,amplitude=10pt},xshift=-4pt,yshift=0pt,thick]
(.9+3*1.95,-2.45) -- (-.6+3*1.95,-2.45);
\node[anchor=center,align=left] at (1.97-1.95,-3.3) {\footnotesize $Y^1|\mathbf{X}^1$};
\node[anchor=center,align=left] at (1.97,-3.3) {\footnotesize $Y^2|\mathbf{X}^2$};
\node[anchor=center,align=left] at (1.97+1.95,-3.3) {\footnotesize $Y^k|\mathbf{X}^k$};
\node[anchor=center,align=left] at (1.97+2*1.95,-3.3) {\footnotesize $Y^K|\mathbf{X}^K$};

\end{scope}

\draw[-,thick] (3,-5) to[bend right=30] (7.5, -5.518);
\draw[-,thick] (3+1.37,-5) to[bend right=30] (7.5, -5.517);
\draw[->,thick] (3+2*1.365,-5) to[bend right=25] (7.6, -5.5);
\draw[-,thick] (3+3*1.363,-5) to[bend right=50] (7.5, -5.52);

\node[align=center] at (8.7,-5.5) {\small\bfseries Invariant \\\small\bfseries across folds};
\node[red!75!black] at(10,-5.5) {\Huge ?};
\end{scope}

\end{tikzpicture}
\caption{
Illustration of a cross-validation scheme across $K$ environments (cf.\ Section~\ref{sec:interventional}).
Environments can correspond to recordings on different days, during different tasks, or on different subjects, and define how the data is split into folds for the cross-validation scheme.
We propose to assess a model
by (a)~leave-one-environment-out cross-validation testing for robust predictive performance on the held-out fold
and
(b)~an invariance-test across environments assessing whether the model is invariant across folds.
The cross-validation scheme (a) is repeated $K$ times, so that each environment acts as a held-out fold once.
Models whose predictive performance does not generalise to held-out data or that are not invariant across environments can be refuted as non-causal.
For linear models, for example, invariance across environments can be assessed by evaluating to which extent regression coefficients and error variances differ across folds (cf.\ Section~\ref{sec:interventional}).
}\label{fig:invariancecv}
\end{figure}

The idea of distributional robustness across
changing background conditions may help us
to falsify
causal hypotheses,
even
when interventional data is difficult to obtain,
and in this sense may guide us towards models that are closer to the causal ground truth.
For this, suppose that the data are obtained in different environments and that we expect
a causal model for $Y$ to yield robust performance across these environments (see Section~\ref{sec:robustcausalmodels}).
Even if we lack targeted interventional data in cognitive neuroscience
and thus cannot test a causal hypothesis directly,
we can test the above implication.
We can test the invariance, for example, using conditional independence tests or
specialised tests for linear models~\parencite{Chow1960}.
We can, as a surrogate,
hold out one environment, train our model on the remaining environments,
and
evaluate how well that model performs on the held-out data
(cf.\ Figure~\ref{fig:invariancecv});
the reasoning is that a non-invariant model may not exhibit robust predictive performance
and instead yield a bad predictive
performance for one or more of the folds.
If a model fails the above
then either
(1) we included the wrong variables,
(2) we have not observed important variables, or
(3) the environment directly affects $Y$.
Tackling (1), we can try to refine our model and search for different variable representations and variable sets that render our model invariant and robust in the post-analysis.
In general, there is no way to recover from (2) and (3), however.

While a model that is not invariant across environments
cannot be the complete causal model (assuming the environments do not act directly on the target variable),
it may still have non-trivial prediction performance and predict better than a simple baseline method in a new, unseen environment.
The usefulness of a model is questionable, however,
if its predictive performance on held-out environments is not
significantly better than a simple baseline.
Conversely, if our model shows robust performance on the held-out data and is invariant across environments, it
has the potential of being a causal model (while it need not be; see Section~\ref{sec:robustcausalmodels} for an example).
Furthermore, a model that satisfies the invariance property is interesting in itself
as it may enable predictions in new, unseen environments.
For this line of argument, it does not suffice to employ a cross-validation scheme that ignores the environment structure and only assesses predictability of the model on data pooled across environments.
Instead, we need to respect the environment structure and assess the distributional robustness of the model across these environments.

For an illustration of the interplay between invariance and predictive performance, consider
a scenario in which
$X_1 \to Y \to H \to X_2$,
where $H$ is unobserved.
Here,
we regard different subjects as different
environments and suppose that (unknown to us) the environment acts on $H$:
One may think of a variable $E$ pointing into $H$.
Let us assume that our study contains two subjects,
one that we use for training and another one that we use as held-out fold.
We compare a model of the form
$\widehat{Y} = f\left(\mathbf{X}_{\pa{Y}}\right) = f(X_1)$
with a model of the form $\widetilde{Y} = g\left(\mathbf{X}\right) = g(X_1,X_2)$.
On a single subject, the latter model including all observed variables has more predictive power than the former model that only includes the causes of $Y$.
The reason is that $X_2$ carries information about
$H$, which can be leveraged to predict $Y$.
As a result, $g(X_1,X_2)$ may predict $Y$ well (and even better than
$f(X_1)$)
on the held-out subject
if it is similar to the training subject
in that the distribution of $H$ does not change between the subjects.
If, however, $H$ was considerably shifted for the held-out subject,
then the performance of predicting $Y$ by $g(X_1,X_2)$ may be considerably impaired.
Indeed, the invariance is violated and we have $E \nindependent Y | X_1, X_2$.
In contrast, the causal parent model $f(X_1)$ may
have worse accuracy on the training subject
but satisfies invariance:
Even if the distribution
of $H$ is different for held-out subjects compared to the training subject,
the predictive performance of the model $f(X_1)$ does not change.
We have $E \independent Y | X_1$.

In practice, we often consider more than two environments.
We hence have
access to several environments when training our model, even if
we leave out one of the environments to test on.
In principle, we can thus already during training distinguish between invariant and non-invariant models.
While some methods have been proposed that explicitly make use of these
different environments during training time (cf.~Section~\ref{sec:methods}),
we regard this as a mainly unexplored but promising area of research.
In Section~\ref{sec:eegloso}, we present a
short
analysis
of classifying
motor imagery conditions on EEG data
that
demonstrates how
leveraging robustness
may yield models that generalise better
to unseen subjects.

In summary,
employing distributional robustness as guiding principle prompts us to reject models as non-causal if they
are not invariant or if they do not generalise better than a simple baseline to unseen environments,
such as sessions, days, neuroimaging modalities, subjects, or other slight variations to the experimental setup.
Models that are distributionally robust and do generalise to unseen environments are not necessarily causal but
satisfy the prerequisites for being
interesting candidate models when it comes to
capturing the underlying causal mechanisms.

\subsubsection{Exemplary proof-of-concept EEG analysis: leave-one-environment-out cross-validation}\label{sec:eegloso}

\begin{figure}
\resizebox{\textwidth}{!}{\includegraphics{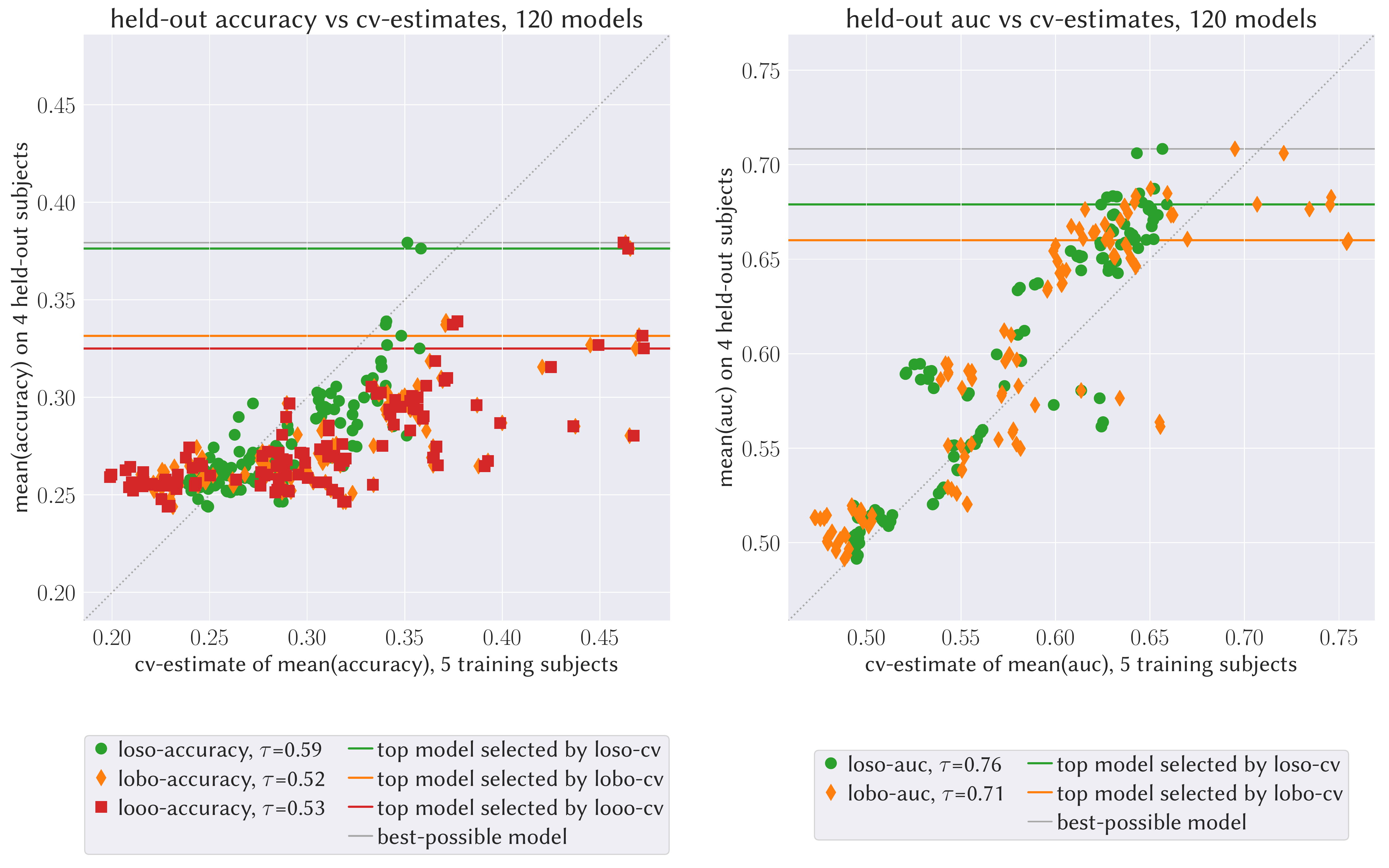}}
\caption{
We compare $120$ models for the  prediction of $4$ motor imagery tasks
that leverage different EEG components, bandpower features in different frequency bands,
and different classifiers.
The left and right panel consider classification accuracy or AUC
averaged over four held-out subjects
as performance measure, respectively.
The leave-one-subject-out (loso) cross-validation accuracy on $5$ training subjects
captures how robust a model is across training subjects.
This leave-one-environment-out cross-validation scheme (see Figure~\ref{fig:invariancecv})
seems indeed able to identify models that
generalise slightly better
to new unseen environments (here the $4$ held-out subjects)
than a comparable $7$-fold leave-one-block-out (lobo)-cv or the leave-one-observation-out (looo) scheme.
This is reflected in the
Kendall's $\tau$ rank correlation
and the scores of the top-ranked models.
All top-ranked
models
outperform
random guessing on the held-out subjects (which corresponds to $25\%$ and $50\%$ in the left and right figure, respectively).
The displacement along the $x$-axis of the lobo- and loso-cv scores
indicates the
previously reported
overestimation of held-out performance when using those cross-validation schemes.
}\label{fig:eeg}
\end{figure}

Here, we illustrate the proposed cross-validation scheme presented in Figure~\ref{fig:invariancecv}
on motor imagery EEG data due to~\textcite{tangermann2012}.
The data consist of EEG recordings of $9$ subjects
performing multiple trials of $4$ different motor imagery tasks.
For each subject $22$-channel EEG recordings at $250$ Hz sampling frequency
are available for $2$ days with $6$ runs of $48$ trials each.
We analysed the publicly available data that is bandpass filtered between $0.5$ and $100$ Hz and $50$ Hz notch filtered.
The data was further preprocessed by re-referencing to common average
reference (car) and projecting onto the orthogonal complement of the null component.
Arguably, the full causal structure in this problem is unkown. Instead of assessing the causal nature of a model directly,
we therefore
evaluate whether
distributional robustness of a model across training subjects
may help to find models that
generalise better to new unseen subjects.

$120$ models were derived on the training data comprising recordings of the first $5$ subjects.
Models relied on $6$ different sets of extracted timeseries components:
the re-referenced EEG channels,
$3$ different sets of $5$ PCA components with varying variance-explained ratios,
and $2$ sets of $5$ coroICA components
using neighbouring covariance pairs and a partition size
of 15 seconds
(cf.\ \textcite{pfister2019coroica} for more details on coroICA).
Signals were bandpass filtered in $4$ different frequency bands
($8-30$, $8-20$, $20-30$, and $58-80$).
For each trial and feature set, bandpower features for classification were obtained
as the log-variance of the bandpass-filtered signal
during seconds $3-6$ of each trial.
For each of the
$6\cdot 4 = 24$ configurations of trial features,
we fitted $5$ different linear discriminant analysis classifiers
without shrinkage,
with automatic shrinkage based on the Ledoit-Wolf lemma,
and shrinkage parameter settings $0.2$, $0.5$, and $0.8$.
These $120$ pipelines were fitted once on the entire training data and
classification accuracies and areas under the receiver operating curve scores
obtained on $4$ held-out subjects
($y$-axes in Figure~\ref{fig:eeg}).
Classifier performance was cross-validated on the training data following
the following three different cross-validation schemes
(cross-validation scores are shown on the $x$-axes in Figure~\ref{fig:eeg}):
\begin{description}
\item[loso-cv]\emph{Leave-one-subject-out cross-validation}
is the proposed cross-validation scheme.
We hold out data corresponding to each training subject once,
fit an LDA classifier on the remaining training data,
and assess the models accuracy on the held-out
training subject.
The average of those cross-validation scores reflects
how robustly each of the $120$ classifier models
performs across environments (here subjects).

\item[lobo-cv]\emph{Leave-one-block-out cross-validation}
is a $7$-fold cross-validation scheme that is similar to the above loso-cv scheme,
where the training data is split into random $7$ blocks of roughly equal size.
Not respecting the environment structure within the training data,
this cross-validation scheme does not capture a models robustness across environments.

\item[looo-cv]\emph{Leave-one-observation-out cross-validation} leaves out a single observation and is equivalent to lobo-cv with a block size of one.
\end{description}

In Figure~\ref{fig:eeg} we display the results of the different cross-validation schemes
and the Kendall's $\tau$
rank correlation between the different cv-scores derived on the training data
and a model's classification performance on the four held-out subjects.
The loso-cv scores correlate more strongly with
held-out model performance and thereby
slightly better resolve the relative model performance.
Considering the held-out performance for the models with top cv-scores,
we observe that selecting models based on the loso-cv score may
indeed select models that tend to perform slightly better on new unseen subjects.
Furthermore,
comparing the displacement
of the model scores
from the diagonal shows that
the loso-cv scheme's estimates are
less-biased
than the lobo and looo cross-validation scores,
when used as an estimate for
the performance on held-out subjects;
this is in line with
\textcite{varoquaux2017assessing}.

\subsection{Robustness and variable constructions}\label{sec:robustconstruction}

Whether
distributional
robustness
holds can depend on
whether
we consider the right variables.
This is shown by the
following example.
Assume
that the target $Y$ is caused by
the two brain signals $X_1$ and $X_2$ via
$$
Y := \alpha X_1 + \beta X_2 + N_Y\text{,}
$$
for some $\alpha\neq0$, $\beta\neq 0$, and noise variable $N_Y$.
Assume further that the environment influences the
covariates $X_1$ and $X_2$
via
$X_1 := X_2 + E + N_1$
and
$X_2 := E + N_2$,
but does not influence $Y$ directly.
Here,
$X_1$ and $X_2$ may represent
neuronal activity in two brain regions that are causal for reaction times while $E$ may
indicate
the time of day or respiratory activity.
We then have the invariance property
$$
E \independent Y \,|\,X_1, X_2.
$$
If, however, we were to construct or---due to limited measurement ability---only be able to observe $\widetilde{X} := X_1 + X_2$,
then whenever $\alpha\neq\beta$
we would find that
$$
E \nindependent Y \,|\,\widetilde{X}.
$$
This conditional dependence is due to
many value pairs for $(X_1,X_2)$
leading to the same value of $\widetilde{X}$:
Given $\widetilde{X}=\widetilde{x}$,
the value of $E$ holds information about
whether say $(X_1,X_2) = (\widetilde{x},0)$
or $(X_1,X_2) = (0, \widetilde{x})$ is more probable
and thus–since $X_1$ and $X_2$ enter $Y$ with different weights–holds information about $Y$;
$E$ and $Y$ are conditionally dependent given $\widetilde{X}$.
Thus, the invariance may break down when aggregating variables in an ill-suited way.
This example is generic in that the same conclusions hold
for
all
assignments
$X_1 := f_1(X_2, E, N_1)$
and
$X_2 := f_2(E, N_2)$,
as long as causal minimality, a weak form of faithfulness, is satisfied \parencite{Spirtes2000}.

Rather than
taking the lack of robustness
as a deficiency, we believe that
this
observation
has the potential to
help us finding the
right variables and granularity
to model our system of interest.
If we are given several environments,
the guiding principle of distributional robustness can
nudge our variable definition and ROI definition towards the construction of variables that are more suitable for causally modelling some cognitive function.
If some ROI activity or some EEG bandpower feature does not
satisfy any invariance
across environments then we may conclude that our variable representation is
misaligned with the underlying causal mechanisms
or that important variables have not been observed
(assuming that the environments do not act on $Y$ directly).

This idea can be illustrated by a
thought experiment that is a
variation of the LDL-HDL example in Section~\ref{sec:challenge2}:
Assume we wish to aggregate multiple voxel-activities
and represent them by the activity of a ROI defined by those voxels.
For example, let us consider the reaction time scenario (Example A, Section~\ref{sec:runningexample})
and voxels $X_1,\dots,X_d$.
Then we may aggregate the voxels $\overline{X}=\sum_{i=1}^d X_i$
to obtain a macro-level model in which we can still sensibly
reason about the effect of an intervention on the treatment variable $T$
onto the distribution of $\overline{X}$, the ROIs average activity.
Yet, the model is in general not causally consistent with the original model.
First, our ROI may be chosen too coarse such that for
$\mathbf{\widehat{x}}\neq\mathbf{\widetilde{x}}\in\mathbb{R}^d$
with $\sum_{i=1}^d \mathbf{\widehat{x}}_i = \sum_{i=1}^d \mathbf{\widetilde{x}}_i = \overline{x}$
the interventional distributions induced by the micro-level model corresponding to setting all $X_i:=\mathbf{\widetilde{x}}_i$ or alternatively to $X_i:=\mathbf{\widehat{x}}_i$
do not coincide—for example, a ROI that effectively captures the global average voxel-activity
cannot resolve whether a higher activity is due to increased reaction-time-driving neuronal entities
or due to some upregulation of other neuronal processes unrelated to the reaction time, such as respiratory activity.
This ROI would be ill-suited for causal reasoning and non-robust as there are two micro-level interventions that imply different distributions on the reaction times while corresponding to the same intervention setting
$\overline{X}:=\overline{x}$
with
$\overline{x} = \sum_{i=1}^d \mathbf{\widehat{x}}_i = \sum_{i=1}^d \mathbf{\widetilde{x}}_i$
in the macro-level model.
Second, our ROI may be defined too fine
grained
such that, for example,
the variable representation does only pick up on the left-hemisphere hub
of a distributed neuronal process relevant for reaction times.
If the neuronal process has different laterality in different subjects,
then predicting the effects of interventions on only the left-hemispherical neuronal activity
cannot be expected to translate to all subjects.
Here, a macro-variable that averages more voxels, say symmetric of both hemispheres, may be more robust to reason about the causes of reaction times
than the more fine grained unilateral ROI.
In this sense, seeking for variable constructions that enable distributionally robust models across subjects,
may nudge us to meaningful causal entities.
The spatially refined and finer resolved cognitive atlas
obtained by \textcite{varoquaux2018atlases},
whose map definition procedure was geared towards an atlas that would be robustness across multiple studies and $196$ different experimental conditions,
may be seen as an indicative manifestation of the above reasoning.

\subsection{Existing methods exploiting robustness}\label{sec:methods}
We now present some existing
methods that explicitly consider the invariance of a model.
While many of these methods are still in their infancy when considering real world applications,
we believe that further development in that area could
play a vital role when tackling causal questions in cognitive neuroscience.

\paragraph{Robust Independent Component Analysis.}
Independent component analysis (ICA) is commonly performed in the analysis of magneto- and electro-electroencephalography (MEG and EEG) data in order to invert the inevitable measurement transformation that leaves us with observations of a linear superposition of underlying cortical (and non-cortical) activity.
The basic ICA model assumes our vector of observed variables $X$ is being generated as $X = AS$ where $A$ is a mixing matrix and $S=[S_1,\dots,S_d]^\top$ is a vector of unobserved mutually independent source signals.
The aim is to find the unmixing matrix $V = A^{-1}$.
If we perform ICA on individual subjects' data separately, the resulting unmixing matrices will often differ between subjects.
This not only hampers the interpretation of the resulting sources as some cortical activity that we can identify across subjects, it also hints---in light of the above discussion---at some unexplained variation that is due to shifts in background conditions between subjects such as different cap positioning or
neuroanatomical variation.
Instead of simply pooling data across subjects, \textcite{pfister2019coroica} propose a methodology that explicitly exploits
the existence of environments, that is, the fact
that EEG samples can be grouped by subjects they were recorded from.
This way, the proposed confounding-robust ICA (coroICA) procedure identifies an unmixing of the signals that generalises to new subjects.
The additional robustness resulted, for their considered example, in improved classification accuracies on held-out subjects and can be viewed as a first-order adjustment for subject specific differences.
The application of ICA procedures to pooled data will generally result in components that do not robustly transfer to new subjects and are thus necessarily variables that do not lend themselves for a causal interpretation.
The coroICA procedure aims to exploit the environments to identify unmixing matrices that generalise across subjects.

\paragraph{%
Causal discovery with exogenous variation.}
Invariant causal prediction,
proposed by
\textcite{Peters2016jrssb},
aims at identifying the
parents of $Y$ within a set of covariates $X_1, \ldots, X_d$.
We have argued
that the true causal model
is invariant across environments, see Equation~\eqref{eq:invariance},
if the data are obtained in different environments and the environment does not
directly influence $Y$.
That is, when enumerating all invariant models by searching through subsets
of $X_1, \ldots, X_d$, one of these subsets must be the set of causal parents of $Y$.
As a result, the intersection
$\widehat{S} = \cap_{S: S \text{invariant}} S$
of all sets of covariates that yield invariant models is
guaranteed to be a subset of the causes $\pa{Y}$ of $Y$. (Here, we define the intersection over the empty index set as the empty set.)
Testing invariance with a hypothesis test to the level $\alpha$, say $\alpha = 0.05$, one obtains that
$\widehat{S}$
is contained in the set of parents of $Y$ with high probability
$$
\mathbf{P}\left(\widehat{S} \subseteq \pa{Y}\right) \geq 1 - \alpha.
$$
Under faithfulness,
the method can be shown to be robust against
violation of the above assumptions.
If the environment
acts on $Y$ directly, for example,
there is no invariant set and
in the presence of hidden variables,
the intersection $\widehat {S}$ of invariant models can still
be shown to be a subset
of the ancestors of $Y$ with large probability.

It is further possible
to model the environment
as a random variable (using an indicator variable, for example),
that is often
called a context variable. One can then
exploit
the background knowledge of its exogeneity
to identify the full causal structure
instead of focussing on identifying the causes of a target variable.
Several approaches have been suggested
\parencite[for example,][]{Spirtes2000, Eaton2007, Zhang-ijcai2017, Mooij++_JMLR_20}.
Often, these methods first identify the target of the intervention and then exploit
known techniques of constraint- or score-based methods. Some of the above methods also make use of
time as a context variable or environment \parencite{Zhang-ijcai2017, Pfister2018jasa}.

\paragraph{Anchor regression.}
We argued above that focusing
on invariance has an advantage
when inferring causal structure from data.
If we are looking for generalisability across environments, however,
focusing solely on invariance
may be too restrictive.
Instead, we may select the most predictive model among all invariant models.
The
idea of
anchor regression
is to explicitly trade off invariance  and predictability~\parencite{rothenhausler2018anchor}.
For a target variable $Y$, predictor variables $\mathbf{X}$, and so-called anchor variables $\mathbf{A} = [A_1, \dots, A_q]^\top$
that represent the different environments
and are normalised to have
unit variance,
the anchor regression coefficients are obtained as solutions to the following minimisation problem
\[
\widehat{b}_\gamma :=\arg\min_{{b}\in\mathbb{R}^d}\ \underbrace{\mathbb{E}\left[ (Y - b^\top \mathbf{X} )^2\right]}_{\text{prediction}} + \gamma \underbrace{\mathbb{E}\left[\| \mathbf{A} (Y - b^\top \mathbf{X})\|^2_{2}\right]}_{\text{invariance}}.
\]
Higher parameters $\gamma$ steer the regression towards more invariant predictions
(converging against the two stage least squares solutions in
identifiable instrumental variable settings).
For $\gamma = 0$ we recover the ordinary least square solution.
The solution $\widehat{b}_\gamma$ can be shown to have the best predictive power under
shift interventions up to a certain strength that depends on~$\gamma$.
As before, the anchor variables can code time, environments, subjects, or other factors, and we thus obtain a regression that is distributionally robust against shifts in those factors.

\subsection{Time series data}\label{sec:furtherprobs}
So far, we have mainly focused on the setting of i.i.d.\ data.
Most of the causal inference literature dealing with time dependency
considers discrete-time models.
This comes with additional complications for causal inference.
For example, there are ongoing efforts to adapt causal inference algorithms and account for sub- or sup-sampling and temporal aggregation~\parencite{danks2013learning,hyttinen2016causal}.
Problems of temporal aggregation relate to Challenge~2 of finding the right variables, which is a conceptual problem in time series models that requires us to clarify our notion of intervention for time-evolving systems~\parencite{rubenstein2017causal}.
When we observe time series with non-stationarities we may consider these as resulting from some unknown shift interventions.
That is, non-stationarities over time may be due to shifts in the background conditions and as such can be understood as shifts in environments analogous to the i.i.d.\ setting.
This way, we may again leverage the idea of distributional robustness for inference on time-evolving systems for which targeted interventional data is scarce.
Extensions of invariant causal prediction to time series data that aim to leverage such variation have been proposed by~\textcite{Christiansen2018, Pfister2018jasa} and the ICA procedure described in Section~\ref{sec:methods} also exploits non-stationarity over time.
SCMs extend to continuous-time models~\parencite{Peters2020},
where the idea to trade off prediction
and invariance has been applied to the problem of inferring
chemical reaction networks~\parencite{Pfister2019pnas}.

A remark is in order if we wish to describe time-evolving systems by one causal summary graph where each time series component is collapsed into one node:
For this to be reasonable, we need to assume a time-homogeneous causal structure.
Furthermore, it requires us to carefully clarify its causal semantics:
While summary graphs can capture the existence of cause-effect relationships, they do in general not correspond to a structural causal model that admits a causal semantics nor enables interventional predictions that are consistent with the underlying time-resolved structural causal model~\parencite{rubenstein2017causal,janzing2018structural}.
That is, the wrong choice of time-agnostic variables and corresponding interventions may be ill-suited to represent the cause-effect relationships of a time-evolving system \parencite[cf.\ Challenge~2 and][]{rubenstein2017causal,janzing2018structural}.

\section{Conclusion and future work}\label{sec:conclusion}

Causal inference in cognitive neuroscience is ambitious.
It is important to continue the open discourse about the many challenges, some of which are mentioned above.
Thanks to the open and critical discourse there is great awareness and caution when interpreting neural correlates~\parencite{rees2002neural}.
Yet, ``FC [functional connectivity] researchers already work within a causal inference framework, whether they realise it or not''~\parencite{reid2019advancing}.

In this article we have
provided our view on
 the numerous obstacles to a causal understanding of cognitive function.
If we, explicitly or often implicitly, ask causal questions, we need to employ causal assumptions and methodology.
We propose to exploit that
causal models using the right variables are distributionally robust.
In particular, we advocate distributional robustness as a guiding principle for causality in cognitive neuroscience.
While causal inference in general and in cognitive neuroscience in particular is a challenging task, we can at least exploit the rational to refute models and variables as non-causal that are frail to shifts in the environment.
This guiding principle does not necessarily identify causal variables nor causal models, but it nudges our search into the right direction away from frail models and non-causal variables.
While we presented first attempts that aim to leverage observations obtained in different environments (cf.~Section~\ref{sec:methods}), this article poses more questions for future research than it answers.

We believe that procedures that exploit environments during training are a promising avenue for future research.
While we saw mildly positive results in our case study, further research needs to show
whether this trend persists in studies with many subjects.
It may be possible to
obtain
improvements when combining
predictive scores on held-out-subjects
with other measures of invariance and robustness.
The development of such methods may be
spurred and guided by field-specific benchmarks (or competitions) that assess models' distributional robustness across a wide range of scenarios, environments, cognitive tasks, and subjects.

When considering robustness or invariance across trainings environments,
the question arises how
the ability to infer causal structure and
the
generalisation performance to unseen environments depend on the
number of training environments.
While first
attempts have been made to theoretically understand that relation
\parencite{rothenhausler2018anchor,
Christiansen2020generalization},
most of the underlying questions are still open.
We believe that an answer
would depend
on the
strength of the interventions,
the sample size,
the complexity of the model class
and, possibly, properties of
the (meta-)distribution
of
environments.

We believe that advancements regarding the errors-in-variable problem may have important implications for cognitive neuroscience.
Nowadays, we can obtain neuroimaging measurements at various spatial and temporal resolutions using, among others, fMRI, MEG and EEG, positron emission tomography, or near-infrared spectroscopy~\parencite{filler2009history,poldrack2018new}.
Yet, all measurement modalities are imperfect and come with different complications.
One general problem is that the observations are corrupted by measurement noise.
The errors-in-variables problem complicates even classical regression techniques where we wish to model $Y \approx \beta X^* + \epsilon$ but only have access to observations of a noise-corrupted $X = X^* + \eta$~\parencite{schennach2016recent}.
This inevitably carries over and hurdles causal inference as the measurement noise spoils conditional independence testing, biases any involved regression steps, and troubles additive noise approaches that aim to exploit noise properties for directing causal edges
and methods testing for invariance. %
First steps addressing these issues in the context of causal discovery have been proposed by
\parencite{Zhang2018uai, Blom++_UAI_18, Scheines2016}.

Summarising, we believe that there is a need for causal models if we aim to understand the neuronal underpinnings of cognitive function.
Only causal models
 equip us with concepts that allow us to explain, describe, predict, manipulate, deal and interact with, and reason about a system
 and that allow us to generalise to new, unseen environments.
A merely associational model suffices to
predict naturally unfolding disease progression, for example.
We need to obtain understanding in form of a causal model
if our goal is to
guide rehabilitation after cognitive impairment or to inform the development of personalised drugs that target specific neuronal populations.
Distributional robustness
and generalisability to unseen environments is an ambitious goal, in particular in biological systems and even more so in complex systems such as the human brain.
Yet, it may be the only and most promising way forward.

\subsection*{Acknowledgments}
The authors thank the anonymous reviewers for
their constructive and helpful feedback on an earlier version of this manuscript.
SW was supported by the Carlsberg Foundation.

\clearpage
\printbibliography

\end{document}